\documentclass[fleqn,10pt]{elsarticle}
\usepackage{lineno}
\usepackage{multirow}
\usepackage{graphicx}
\usepackage{epstopdf}
\usepackage{mathrsfs}
\usepackage{amssymb}
\usepackage{amsmath}
\usepackage{color}
\usepackage{longtable}
\usepackage{lscape}
\usepackage{multicol}
\allowdisplaybreaks
\modulolinenumbers[5]

\journal{Journal of Theoretical Biology}

\newcommand{\bm}[1]{\boldsymbol{#1}}

\begin{document}
\title{Dynamics of cell-type transition mediated by epigenetic modifications}

\author[a]{Rongsheng Huang}
\address[a]{School of Science, Jimei University, Xiamen Fujian 361021, China}

\author[b]{Qiaojun Situ}
\address[b]{Zhou Pei-Yuan Center for Applied Mathematics, Tsinghua University, Beijing 100084, China}

\author[c]{Jinzhi Lei \corref{correspondingauthor}}
\address[c]{School of Mathematical Sciences, Center for Applied Mathematics, Tiangong University, Tianjin, 300387, China}
\cortext[correspondingauthor]{Corresponding author: jzlei@tiangong.edu.cn}

\date{\today} 

\begin{abstract}
Maintaining tissue homeostasis requires appropriate regulation of stem cell differentiation. The Waddington landscape posits that gene circuits in a cell form a potential landscape of different cell types, wherein cells follow attractors of the probability landscape to develop into distinct cell types. However, how adult stem cells achieve a delicate balance between self-renewal and differentiation remains unclear. We propose that random inheritance of epigenetic states plays a pivotal role in stem cell differentiation and present a hybrid model of stem cell differentiation induced by epigenetic modifications. Our comprehensive model integrates gene regulation networks, epigenetic state inheritance, and cell regeneration, encompassing multi-scale dynamics ranging from transcription regulation to cell population. Through model simulations, we demonstrate that random inheritance of epigenetic states during cell divisions can spontaneously induce cell differentiation, dedifferentiation, and transdifferentiation. Furthermore, we investigate the influences of interfering with epigenetic modifications and introducing additional transcription factors on the probabilities of dedifferentiation and transdifferentiation, revealing the underlying mechanism of cell reprogramming. This \textit{in silico} model provides valuable insights into the intricate mechanism governing stem cell differentiation and cell reprogramming and offers a promising path to enhance the field of regenerative medicine.  
\end{abstract} 


\maketitle 

\section{Introduction}
Adult stem cells play a vital role in maintaining tissue homeostasis by replenishing dying cells and regenerating damaged tissues through controlled self-renewal and differentiation\cite{Clevers2015}. Understanding the mechanism underlying cell fate decisions and the regulation of self-renewal and differentiation in stem cell biology is of utmost significance.

Waddington's epigenetic landscape is a fundamental concept in comprehending cell fate decisions and cell differentiation\cite{Waddington2012}. The landscape analogy visualizes a cell as a ball rolling on a mountain, with valleys representing stable cell phenotype  {and ridges signifying cell fate choice leading to} new phenotypes.  {While} Waddington's epigenetic landscape provides an intuitive understanding of the biological process, 
  {the mechanisms driving cellular development and reprogramming remain elusive\cite{Ferrell2012}.}  {Specifically, how gene circuits define Waddington's epigenetic landscape in a multicellular system, the driving force behind cell fate decision, and how adult stem cell systems balance self-renewal and differentiation pose intriguing questions.}  

 {Gene circuits consisting} of two transcription factors, such as PU.1-GATA1 or SOX2-OCT4,  {have been extensively} studied  {as they are} associated with cell fate decisions\cite{Jolly2015a,Wang2011,Iovino2011,Zhou2011}.  {These circuits exhibit three stable equilibrium points corresponding to stem cells and two differentiated cell fates, validating the concepts} of Waddington landscape.  {The regulatory interplay between these transcription factors determines cell fate. Despite such discoveries, the driving force triggering cell-type switches remains a topic of debate, with stochastic fluctuation, gene regulation, cell-cell communications, and artificial induction being proposed as potential mechanisms\cite{Rommelfanger2021,Xu2014,Losick2008,Cobaleda2007,Zhou2008,Zhou2008a,Yamanaka2012,Amabile2009,Shu2013}.}  {Remarkably, self-renewal and differentiation of stem cells maintain a state of dynamic equilibrium even in the face of random environmental changes. This highlights the importance of regulating the transitions between cell types during stem cell regeneration to ensure reliable tissue function.}

 {Recent advancements in} single-cell sequencing techniques have  {illuminated} cell heterogeneity,  {revealing macro-heterogeneities in gene expression among cells with different phenotypes and even microscopic heterogeneity within cells of the same phenotype}\cite{Angermueller2016,Pollen2014,Patel2014,Huang2009,Bjorklund2016a,Johnson2015,Shalek2013}. Epigenetic regulation,  {encompassing histone modification and DNA methylation, has emerged as a key player in} cellular heterogeneity and phenotype switching\cite{Lee2014,Bagci2013,Zhang2016a,Feinberg2023}.

Histones are structural chromosome proteins, including H1 and H3, H4, H2A, and H2B  {subunits}. There  {are various} forms and multiple functions of histone modifications  {that} regulate gene expression.  {For instance,} trimethylation of lysine 4 on histone H3 protein subunit (H3K4me3) and H4K12 acetylation can promote gene expression,  {while} ubiquitination, such as ub-H2A, can inhibit gene expression\cite{Rossetto2012,Plass2013}. Histone modifications  {are} heritable during cell division. The parental modifications  {are} recognized by a binding protein or a reading protein,  {which then recruits} a chromatin modifier or writer protein to alter the histone modifications\cite{Probst2009}. 

DNA methylation is a  {process where} methyl($\mathrm{CH}_3$) group  {is added} to DNA,  {which can alter} the function of genes and  {impact} gene expression.  {Both histone modifications and DNA methylation inheritance} are semi-conservative but inevitably accompanied by natural changes during each generation, which may  {lead to} cell heterogeneity and plasticity\cite{Easwaran2014}. 

 {In this study}, we  {present} a hybrid model of stem cell regeneration  {that incorporates} cell phenotype changes induced by epigenetic modifications. The model integrates a gene regulation network, epigenetic state inheritance, and cell regeneration,  {enabling} multi-scale dynamics from transcription regulation to cell population. Through model simulations, we  {explore} how random inheritance of epigenetic states during cell division can  {automatically} induce cell differentiation, dedifferentiation, and transdifferentiation.   {Our \textit{in silico} model offers valuable insights into} the mechanism of stem cell differentiation and cell reprogramming. 

\section{Hybrid model of stem cell differentiation}
\label{model}

\begin{figure}[htbp]
\centering
\includegraphics[width=10cm]{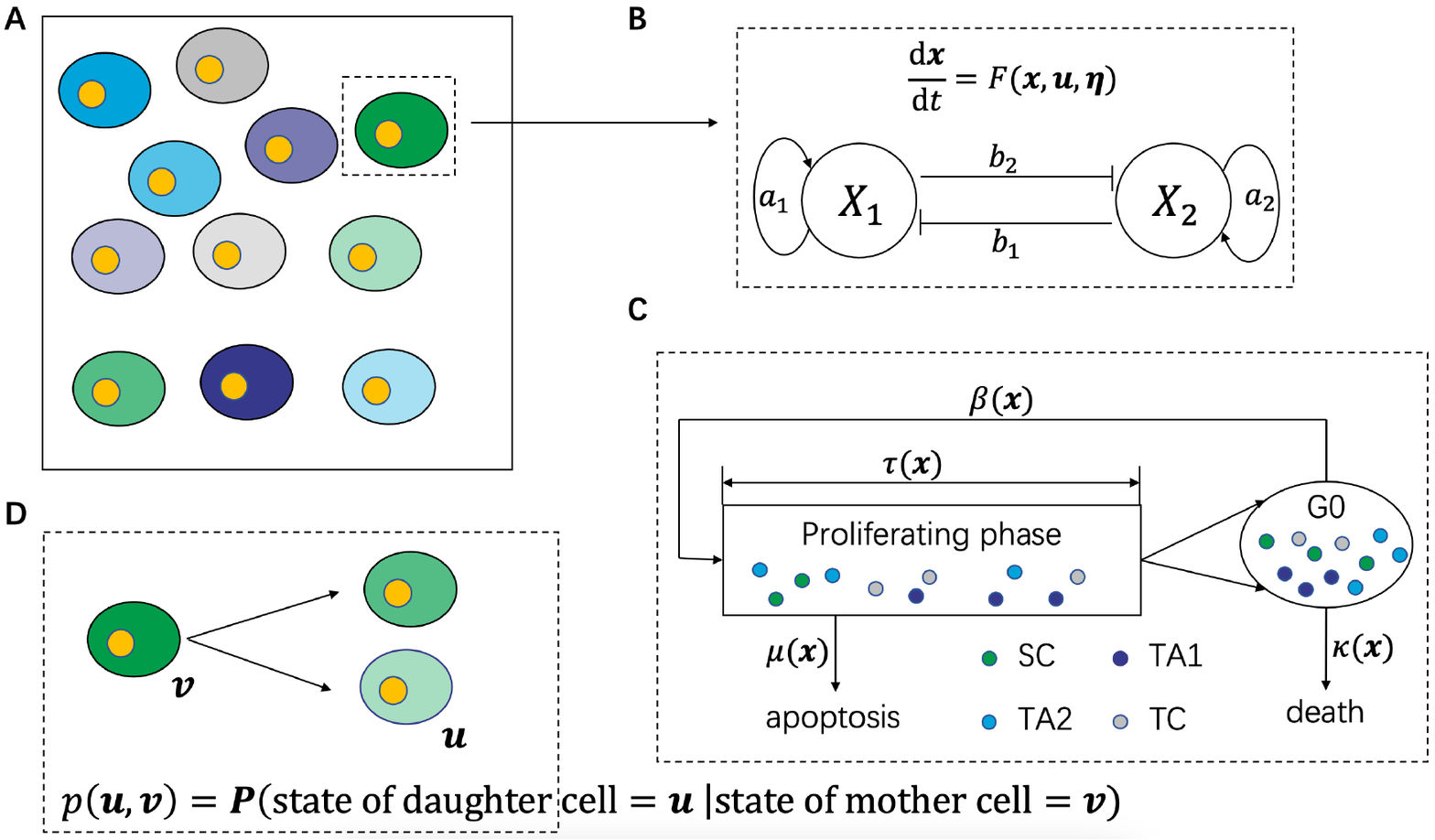}
\caption{Illustration of the hybrid model. (A) Individual-based model of a multi-cellular system. (B) Dynamic system of gene circuit motif. (C) G0 cell cycle model of cell regeneration. (D) Stochastic inheritance of epigenetic state during cell divisions. }
\label{fig:model}
\end{figure}

The hybrid model established in this study is illustrated in Figure \ref{fig:model}.  {It consists of} individual-based modeling of a multi-cellular system (Fig. \ref{fig:model}A), a gene regulation network (GRN) dynamics of two genes that self-activate and repress each other (Fig. \ref{fig:model}B), a G0 cell cycle model  {for cell regeneration} (Fig. \ref{fig:model}C), and stochastic inheritance of epigenetic states during cell divisions (Fig. \ref{fig:model}D).  {The three types of dynamics are coupled with each other through the cell cycle processes and cell-type transitions. The} detailed formulations of the model are given below.   

\subsection{Gene regulation network}
To investigate how epigenetic modifications can drive cell lineage commitment, we consider a GRN  {consisting} of two master transcription factors (TFs) $X_1$ and $X_2$,  {which} self-activate and repress each other (Fig. \ref{fig:model}B). This gene network frequently  {appears} in many cell-fate decision-making  {systems} and has been extensively studied\cite{Zhou2011,Niwa2005,Schaffer2010,Ralston2005,Orkin2008,Loh2008,Wang2010a}.   {Mathematically, the gene network can be modeled using two different approaches: additive and multiplicative regulations, where the production rates from positive and negative feedbacks are either added or multiplied, respectively\cite{Jolly2015a,Xu2014}. This study adopts the additive modeling approach, following the model described in \cite{Xu2014}.  However, it is essential to note that the model formulations and results presented in this study are also consistent with multiplicative feedback regulations.} 

Let $x_1$ and $x_2$ represent the expression level (protein concentration) of genes $X_1$ and $X_2$, respectively. The gene expression dynamics within one cell cycle are modeled with the following ordinary differential equations:
 \begin{equation}
 \label{eq:1}
 \left\{
\begin{aligned}
&\dfrac{\mathrm{d}x_1}{\mathrm{d}t} = a_1\left(\rho_1+(1-\rho_1)\dfrac{x_1^{n}}{s_1^n+x_1^{n}}\right)+b_1\dfrac{s_2^{n}}{s_2^n+x_2^{n}}-k_1x_1,\\
&\dfrac{\mathrm{d}x_2}{\mathrm{d}t} = a_2\left(\rho_2+(1-\rho_2)\dfrac{x_2^{n}}{s_2^n+x_2^{n}}\right)+b_2\dfrac{s_1^{n}}{s_1^n+x_1^{n}}-k_2x_2,
\end{aligned}
\right.
\end{equation}
where $a_1$, $a_2$, $\rho_1$, $\rho_2$, $s_1$, $s_2$, $n$, $b_1$, $b_2$, $k_1$, and $k_2$ are non-negative parameters. The parameters $a_1$ and $a_2$ denote the maximum expression rates of the self-activation of the two genes,   {while} $\rho_1$ and $\rho_2$ ($0\leq \rho_i \leq 1$) represent the ratios between the basal level to the maximum level of the regulation of each gene. The parameters $b_1$ and $b_2$ denote the basal expression rates of the two genes without repression. The parameters $s_1$ and $s_2$ represent the half-effective concentration of the two proteins $X_1$ and $X_2$, respectively, in the transcription regulation,  {and $n$ represents the corresponding Hill coefficient.} The degradation rates of the two proteins are represented by $k_1$ and $k_2$, respectively. 

To  {address the influence} of extrinsic noise perturbation, we introduced stochastic fluctuations to $a_1$ and $a_2$, resulting in the following expressions:
$$
 {a_i(\eta_i)=\alpha_i e^{\sigma_i\eta_i-{\sigma_i^2}/{2}},\quad i=1, 2,}
$$
 {where $\alpha_1$ and $\alpha_2$ are positive parameters to represent the average expression rates}. Here, $\sigma_1$ and $\sigma_2$ represent the intensities of the noise perturbations, and $\eta_1$ and $\eta_2$ are color noises defined by the Ornstein-Uhlenbeck processes:
 \begin{equation}
 \label{eq:4}
 {
\mathrm{d}\eta_i=-(\eta_i/\zeta_i)\mathrm{d}t+\sqrt{2/\zeta_i}\mathrm{d}W_i(t),\quad i=1, 2,}
\end{equation}
where $W_1(t)$ and $W_2(t)$ are independent Wiener process, and $\zeta_1$ and $\zeta_2$ are relaxation coefficients. 
 {In the stationary state, we have  {$\mathrm{E}(a_i(\eta_i)) = \alpha_i$ and $\mathrm{E}\left( \eta_i(t_1)\eta_i(t_2)\right) = e^{-|t_1 - t_2|/\zeta_i}$ ($i=1,2$), }where $\mathrm{E}(\cdot)$ represents the mathematical expectation.}

To  {further account for} the effects of epigenetic modification  {on} gene regulation dynamics,  {it is known} that epigenetic regulations, such as histone modification or DNA methylation, can interfere with chromatin structure and alter the basal expression rates.  {Therefore, we introduced the assumption} that the expression levels $a_1$ and $a_2$  {are dependent on} the epigenetic modification state of the two genes, denoted by $u_1$ and $u_2$, respectively.  {Here, by the epigenetic state, we mean the fractions of marked nucleosomes or methylated CpG sites in a DNA segment of interest. Hence, the epigenetic state $\bm{u} = (u_1, u_2)$ is assumed to lie in the domain $\Omega = [0,1]\times [0,1]$. Additionally, we considered that the epigenetic states $u_1$ and $u_2$ undergo random changes only during cell division, which is discussed in more detail below.}

 {Since epigenetic states primarily affect chromatin structure, they might influence the chemical potential required} to initiate the transcription process. Thus,  {along with the extrinsic noise perturbations,} we can express the expression rates $a_1$ and $a_2$ as follows
  \begin{equation}
  \label{eq:y_to_a_2}
 {a_i(u_i, \eta_i) = \alpha_i e^{\lambda_i u_i} e^{\sigma_i\eta_i-{\sigma_i^2}/{2}},\quad i=1,2,}
\end{equation}
where $\lambda_i (i=1,2)$ represent the  {impact of} the epigenetic modification states  {on the expression levels}. Specifically, $\lambda_i > 0$ indicates an epigenetic modification that  {enhances} the strength of self-activation, while $\lambda_i<0$ indicates a modification that reduces this strength. 

  {
The expression state $\bm{x}= (x_1, x_2)$ depends on the epigenetic state $\bm{u}= (u_1, u_2)$ within one cell cycle through equations \eqref{eq:1}-\eqref{eq:y_to_a_2}.} With properly selected parameter values, the model exhibits three stable steady states that correspond to three cell types (see Fig. \ref{fig:res1_1} below). For instance, the PU.1-GATA1 motif is involved in the gene circuit that determines the cell fate of erythroid/megakaryocyte lineages from granulocyte/monocyte lineages. PU.1 and GATA1 are expressed in the precursor cells (PC).  {Specifically, the granulocyte/monocyte lineage (GMC) cells exhibit low PU.1 expression and high GATA1 expression, while erythroid/megakaryocyte lineage (EMC) cells show high GATA1 expression and low PU.1 expression. For the sake of generality, we refer to the three cell types as stem cell (SC), transit-amplifying cell 1 (TA1), and transit-amplifying cell 2 (TA2). Additionally, we use transition cells (TC) for cells not classified as SC, TA1, or TA2. Mathematically, the phenotypes are defined from $\bm{x} = (x_1, x_2)$ as (see Section \ref{sec:3.1} below):
\begin{equation}
\label{eq:phenotype}
\mbox{phenotype} = 
\left\{
\begin{aligned}
\mbox{SC}:\quad  &\mbox{both}\ X_1\ \mbox{and}\ X_2\ \mbox{are medium expression}, \\
\mbox{TA1}:\quad  &X_1\ \mbox{is high expression and}\ X_2\ \mbox{is low expression}, \\
\mbox{TA2}:\quad  &X_1\ \mbox{is low expression and}\ X_2\ \mbox{is high expression},\\
 {\mbox{TC}}:\quad  &\mbox{otherwise}.\\
\end{aligned}
\right.
\end{equation}}

 {We note that both positive and negative feedback parameters $a_i$ and $b_i$ are subjected to noise perturbation and epigenetic regulations. Here, we only consider the changes in the positive feedback parameters for the simplicity of model introduction and discussions. The methods and results in this study can be extended to general cases with modifications in both feedback parameters. }

 {Our model formulates random perturbations to expression rate by multiplying the log-normal distribution random number defined by an Ornstein-Uhlenbeck process. This type of formulation was introduced in our previous studies\cite{Lei2009} and differs from the conventional Langevin approach.  Biologically, gene expression rates have been observed to follow a log-normal distribution rather than a normal distribution\cite{Austin2006}. Mathematically, using the expression in equation \eqref{eq:y_to_a_2} can prevent the possible negative expression rate, which can be problematic when adding a Gaussian perturbation to the expression rate $a_i$. This formulation is biologically more appropriate for modeling extrinsic random perturbations.}

\subsection{G0 cell cycle model}
To incorporate the above gene regulation network dynamics with cell division, we referred to the G0 cell cycle model of heterogeneous cell regeneration
\cite{Lei2014a,Lei2020a,Lei2020}. In this model, we only consider cells with the ability  {to undergo} cell cycling, and each cell has different rates of proliferation and cell death dependent on its  {cell phenotypic (SC, TA1, TA2, or TC)}.  {Cells that have lost} the ability to undergo cell cycling were not considered and were removed from the system. The cycling cells are classified into resting (G0) or proliferating phases. Resting phase cells can either re-enter the proliferating phase with a rate $\beta$ or be removed from the resting phase with a rate $\kappa$ due to cell death or senescence. Proliferating phase cells can either  {randomly exit} with a rate $\mu$ due to apoptosis or divide into two daughter cells  {after a time $\tau$ following} entry into the proliferative compartment (Fig. \ref{fig:model}C).  { The kinetic rates of each cell, including the proliferation rate $\beta$, the removal rate $\kappa$, the apoptosis rate $\mu$, and the proliferation duration $\tau$ depend on the corresponding cell phenotype.}

The   {SC, TA1, TA2, and TC} cells differ in their regulation of cell proliferation. For SCs, the self-renewal ability is biologically associated with microenvironmental conditions and intracellular signaling pathways \cite{Moustakas2002,Yang2010}. Despite the complexity of signaling pathways, the phenomenological formulation of the Hill function dependence can be derived from simple assumptions regarding the interactions between signaling molecules and receptors \cite{Lei2020a,Bernard2003}, and is given by
$$
\beta_{\text{SC}} = \beta_0 \dfrac{\theta}{\theta + Q(t)},
$$
where   {$Q(t)$} represents the number of SC at time $t$, $\beta_0$ represents the maximum proliferation rate, and $\theta$ is a constant for the half-effective cell number.   {We define the removal rate $\kappa$, the apoptosis rate $\mu$, and the proliferation duration $\tau$ of stem cells as:}
$$
 {\kappa_{\text{SC}} = \kappa_0, \quad \mu_{\text{SC}} = \mu_0, \quad \tau_{\text{SC}} = \tau_0.}
$$
 {For TA1 or TA2 cells, we assumed that they have unconstrained proliferation rates, higher removal rates, and shortened proliferation durations than stem cells.} Moreover, each  {TA1 or TA2 cell} has a limited ability of cell divisions, \textit{i.e.}, a  {TA1 or TA2 cell will be removed from the system} when it reaches the maximum cell division times (here we set it as $15$). Hence, we have
 $$
 \beta_{\text{TA1}}= \beta_{\text{TA2}}= \beta_0,\quad  \mu_{\text{TA1}} =\mu_{\text{TA2}}=\mu_0,\quad \tau_{\text{TA1}}= \tau_{\text{TA2}}=\tau_0/2,
 $$
$$
\kappa_{\text{TA1}}=\kappa_{\text{TA2}}=\left\{\begin{array}{ll}
2\kappa_0, & \mbox{divisions} < 15,\\
+\infty, & \mbox{divisions} \geq 15,
\end{array}
\right.
$$

The transition state is usually very short between cell divisions; hence, cell proliferation and apoptosis are not considered for TCs. Thus, we set 
 $$
 \beta_{\text{TC}}=\mu_{\text{TC}} =\tau_{\text{TC}}=\kappa_{\text{TC}}=0.
 $$
Table \ref{tab:1} summarizes  {the kinetic rates for different phenotypes}. 

\begin{table}[hbtp]
\centering
\caption{  {The kinetic rates for different phenotype}}
\begin{tabular}{c|cccc}
\hline
\hline
 {cell phenotype}  &   {$\beta$} &   {$\kappa$ }&   {$\mu$} &   {$\tau$}\\
\hline
SC &   {$\beta_0 \theta/(\theta + Q)$} & $ \kappa_0$  & $\mu_0$ & $\tau_0$ \\
TA1  & $\beta_0$ &  {${2 \kappa_0}$ or $+\infty$} & $\mu_0$ & $\tau_0/2$\\
TA2  & $\beta_0$ &   {${2 \kappa_0}$ or $+\infty$} & $\mu_0$ & $\tau_0/2$\\ 
  {TC} & {$0$} &  {$0$} &  {$0$} &  {$0$}\\
\hline
\hline
\end{tabular}
\label{tab:1}
\end{table}

From the above model description, given the epigenetic state $\bm{u} = (u_1, u_2)$ of each cell, the gene expression state $\bm{x} = (x_1, x_2)$ dynamically evolves according to the stochastic differential equations \eqref{eq:1}-\eqref{eq:y_to_a_2}. Accordingly, the cell phenotype and the kinetic rates   {$\beta$},   {$\kappa$},   {$\mu$} and   {$\tau$} can change during a cell cycle. In stochastic simulations, we model each cell's random proliferation, apoptosis, and cell-type switches in a multiple-cell system. Each cell has its cell state and randomly undergoes proliferation, apoptosis, and death with a probability depending on the cell state. Finally, when a cell undergoes mitosis, the cell divides into two cells, and the epigenetic states of the two daughter cells are calculated based on the inheritance probability functions below.

 {The effect of cell volume change was not considered in the model. Cell growth and volume changes are important biological processes and play important roles in cell-fate decisions\cite{Doncic2011,Ginzberg2015}. Moreover, biological mechanisms controlling cell growth and division are complicated and unclear for mammalian cells\cite{Kafri2013,Cadart2018,Zatulovskiy2020}. In this study, to avoid the complexity and be more focused on the epigenetic regulations, we assumed that the volume is unchanged during cell cycling so that cell growth does not affect the protein concentration.}

\subsection{Stochastic inheritance of epigenetic states}

Histone modifications and DNA methylations in the daughter cells are reconstructed during cell division based on those in the mother cells. The epigenetic states of the daughter cells usually differ from those of their mother cells, and there is a random transition of the epigenetic states during cell division\cite{Probst2009}.  Moreover, the molecules (proteins and mRNAs) in the mother cells undergo random partition during mitosis,  {leading to their reallocation} to the two daughter cells. Therefore, after mitosis, the epigenetic state $\bm{u}$ and the expression level (protein concentration) $\bm{x}$ of the model \eqref{eq:1}-\eqref{eq:y_to_a_2} for each newborn cell are reset. Here, for simplicity, we assumed symmetry division, meaning that the gene expression state $\bm{x}$ of the two daughter cells is the same as that of the mother cell. However, the epigenetic state $\bm{u}$ may undergo random transitions during cell division. 

To model the stochastic inheritance of epigenetic states during cell division,  {we assumed that the epigenetic states of the two daughter cells are independent of each other, and} we introduced an inheritance function $p(\bm{u}, \bm{v})$ to represent the conditional probability that a daughter cell of state $\bm{u}$ comes from a mother cell of state $\bm{v}$ after cell division. In other words,  
$$
p(\bm{u}, \bm{v}) = P(\mbox{state of daughter cell} = \bm{u}\ | \ \mbox{state of mother cell} = \bm{v}). 
$$
The inheritance function represents cell plasticity in each cell cycle, while the detailed biochemical processes of cell division are ignored. It is obvious to have
$$
\int_\Omega p(\bm{u}, \bm{v}) \mathrm{d} \bm{u} = 1,\quad \forall \bm{v}\in \Omega.
$$ 
Biologically, the exact formulation of the inheritance function $p(\bm{u}, \bm{v})$ is difficult to determine, as it depends on the complex biochemical reactions during the cell division process. Nevertheless, while we consider $p(\bm{u}, \bm{v})$ as a conditional probability density, we focus on the epigenetic state before and after cell division and omit the intermediate complex process.  {This allows us to introduce a} phenomenological function through numerical simulation based on a computational model of histone modification inheritance\cite{Lei2020a,Huang2018,Huang2019}. 

 {The states $u_1$ and $u_2$ represent the epigenetic states at two DNA segments corresponding to the gene $X_1$ and $X_2$, respectively, and we assumed that they vary independently during cell division. If otherwise, we need further assumptions about their interdependence.} Thus, we have
\begin{equation}
p(\bm{u}, \bm{v}) = p_1(u_1, \bm{v}) p_2(u_2, \bm{v}),
\end{equation}
where $p_i(u_i, \bm{v})$ represents the transition function of $u_i$, given the state $\bm{v}$ of the mother cell. According to \cite{Huang2018,Huang2019}, the normalized nucleosome modification level of daughter cells can be described by a random beta-distribution number dependent on the mother cell. Thus, we can write the inheritance function $p_i(u_i, \bm{v})$ through the density function of beta-distribution as 
\begin{equation}
p_i(u_i,\bm{v})=\frac{u_i^{g_i(\bm{v})-1}(1-u_i)^{h_i(\bm{v})-1}}{B(g_i(\bm{v}),h_i(\bm{v}))},\quad  B(g,h)=\frac{\Gamma(g)\Gamma(h)}{\Gamma(g+h)},
\end{equation}
where $\Gamma(z)$ is the gamma function,  $g_i(\bm{v})$ and $h_i(\bm{v})$ are shape parameters that depend on the epigenetic state of the mother cell. We assumed that the conditional expectation and conditional variance of $u_i$ are (given the state $\bm{v}$)
$$
\mathrm{E}(u_i | \bm{v})=\phi_i(\bm{v}),\quad \mathrm{Var}(u_i | \bm{v})=\frac{1}{1+\psi_i(\bm{v})}\phi_i(\bm{v})(1-\phi_i(\bm{v})),
$$
where the shape parameters can be expressed as 
$$
g_i(\bm{v})=\psi_i(\bm{v})\phi_i(\bm{v}),\quad h_i(\bm{v})=\psi_i(\bm{v})(1-\phi_i(\bm{v})).
$$
Here, we note that $\phi_i(\bm{v})$ and $\psi_i(\bm{v})$ always satisfy
$$
0<\phi_i(\bm{v})<1,\quad \psi_i(\bm{v})>0.
$$
Hence, the inheritance function $p(\bm{u},\bm{v})$ can be determined by the predefined conditional expectation and conditional variance, \textit{i.e.}, the functions $\phi_i(\bm{u})$ and $\psi_i(\bm{u})$.  {Here, we assumed that $\psi_i(\bm{v})$ remains constant, while $\phi_i(\bm{v})$ increases with $v_i$ and is expressed by a Hill function as:}
\begin{equation}
\label{eq:epiphi}
\psi_i(\bm{v})=m_{0},\quad \phi_i(\bm{v})=m_{1}+m_{2}\frac{(m_{3}v_i)^{m_{4}}}{1+(m_{3}v_i)^{m_{4}}},\quad \bm{v} = (v_1, v_2),\ i=1,2,
\end{equation}
where $m_{j} (j=0,1, 2, 3, 4)$ are all positive parameters. 

From the above formulation, given the functions $\psi_i(\bm{v})$ and $\phi_i(\bm{v})$, the inheritance function $p(\bm{u}, \bm{v})$ can be expressed as a density function of   {beta-distribution} random numbers. 

\subsection{Numerical scheme}

The proposed hybrid model describes the dynamics of gene regulation networks and cell-type switches of individual cells in a multicellular system. Here, we present an individual-based numerical scheme aimed at simulating the dynamics of each cell in the system. 
\begin{itemize}
\item \textbf{Initialization}: Set the time $t = 0$, the cell number $N$, and the state of all cells $\Sigma = \{[C_i(\bm{x}_i, \bm{u}_i)]_{i=1}^N\}$. Determine the phenotype  (SC, TA1,  TA2,  {or TC}) and proliferation state (resting or proliferating phase) of each cell, and compute the count $Q$ of stem cells. All cells are initially set as stem cells in the resting phase. Accordingly, set the division number of each cell as $\mathrm{div}_i = 0$, and the corresponding age at the proliferating phase (starting from the entry of proliferating phase) as $a_i = 0$.  
\item \textbf{Iteration}: \textbf{for} $t$ from $0$ to $T$ with a time step of $\Delta t$, \textbf{do}:
\begin{itemize}
\item[] \textbf{for} each cell in $\Sigma$, \textbf{do}:
\begin{itemize}
\item Numerically solve equations \eqref{eq:1}-\eqref{eq:y_to_a_2} for a step $\Delta t$  {and} update the expression state $\bm{x}$.  If the cell is in the resting phase, update the phenotype of the cell based on the state $\bm{x}$.
\item Calculate the proliferation rate $\beta$, the apoptosis rate $\mu$, the terminate differentiation rate $\kappa$, and the proliferation duration $\tau$.
\item Determine the cell fate during the time interval $(t, t + \Delta t)$:
\begin{itemize}
\item When the cell is at the resting phase, remove the cell from the simulating pool with a probability $\kappa \Delta t$, or enter the proliferating phase with a probability $\beta \Delta t$. If the cell enters the proliferating phase, set the age $a_i = 0$, and if the cell is a TA cell, set $\mathrm{div}_i = \mathrm{div}_i + 1$.  
\item When the cell is in the proliferating phase, if the age $a_i < \tau$, the cell is either removed (through apoptosis) with a probability $\mu \Delta t$, or remains unchanged and $a_i = a_i + \Delta t$. If the age $a_i \geq \tau$, the cell undergoes mitosis and divides into two cells. When mitosis occurs, set the state of two daughter cells following the rules below: set the age $a_i = 0$; set the epigenetic state $\bm{u}$ of each daughter cell according to the inheritance probability function $p(\bm{u}, \bm{v})$; set the gene expression state $\bm{x}$ as that of the mother cell.
\item After mitosis, check the division number of TA cells. If the division number $\mathrm{div}_i$ of a TA cell is larger than the maximum value, remove the cell from the simulating pool.
\end{itemize}
\end{itemize}
\textbf{end for}
\item[] \textbf{Update} the system $\Sigma$ with the cell number, epigenetic and gene expression states of all surviving cells, and the ages of the proliferating phase cells, and set $t = t + \Delta t$.
\end{itemize}
\textbf{end for}
\end{itemize}

 {In model simulations, we set the time step $\Delta t = 0.25$h.} The numerical scheme can be implemented by the object-oriented programming language \verb|C++|\footnote{ {The source codes are available at \texttt{https://github.com/jinzhilei/Cell-type-transition}.}}.

Table \ref{tab:para} lists the parameter values used in the current study. Here, the parameters illustrate the general process of phenotype switches due to epigenetic modifications along the cell regeneration process  {and are not specific to a particular type of cells}.   {The parameters for the gene regulatory network were selected so that the corresponding gene network has multiple stable states for different cell types. The parameters for the epigenetic regulation were chosen so that cell-type transition can occur with a reasonable frequency. The parameter for cell regeneration (rate of proliferation, differentiation, and cell death) were taken per the biological restriction for the kinetic rates.}  Additionally, we only considered symmetric parameters, which means that $b_1 = b_2, k_1 = k_2, s_1 = s_2, \rho_1 = \rho_2, {\alpha_1=\alpha_2}$. Nevertheless, the main results are insensitive to these parameter values.

\begin{table}[hbtp]
\centering
\caption{Parameter values for model simulation}
\label{tab:para}
\begin{tabular}{lp{6.5cm}ll}
 \hline
 \hline
Parameter & Description &Value & Unit$^{(a)}$\\
  \hline
$b_1,b_2$ &Strengths of the mutual inhibition & $1$ & ${\mathrm{AU}\times \mathrm{h}^{-1}}$\\
$k_1,k_2$ &Degradation rates of the proteins & $1$ & $\mathrm{AU}\times \mathrm{h}^{-1}$\\
$s_1,s_2$ &$50\%$ effective concentration of the feedback loops & $0.5$ & $\mathrm{AU}$\\
$\rho_1,\rho_2$ & The ratio between the basal level to the maximum level of the regulation of each gene & $0.1$ & -\\
$\sigma_1,\sigma_2$ &Intensity of the noise& $0.05$ & - \\
$\zeta_1,\zeta_2$ &Relaxation coefficient of the Ornstein-Uhlenbeck process & $1$ & $\mathrm{h}^{-1}$\\
$\alpha_1,\alpha_2$ &Expression rate of each gene & $0.4$ & $\mathrm{AU}\times \mathrm{h}^{-1}$\\
$\lambda_1,\lambda_2$ &Impact coefficient of epigenetic modifications & $1.9$ & - \\
$n$ & Hill coefficient & 4 & -\\
\hline
$m_0$ &constant & $60$ & -\\
$m_1$ &constant & $0.08$ & -\\
$m_2$ &constant & $1.0$ & -\\
$m_3$ &constant & $2.0$ & -\\
$m_4$ &Hill coefficient & $2.0$ & -\\
\hline
$\theta$ &Constant for the half-effective cell number for SC population control& 200 & cells\\
$\beta_0$ &Maximum proliferation rate& $0.04$ & $\mathrm{h}^{-1}$\\
$\kappa_0$ &The rate of removing cells out of the resting phase & $0.01$ & $\mathrm{h}^{-1}$\\
$\mu_0$ &The apoptosis rate of cells in the proliferating phase & $0.0002$ & $\mathrm{h}^{-1}$\\
$\tau_0$ &Proliferating phase time & $4$ & h\\
\hline
\hline
\end{tabular}

\begin{minipage}{8cm}
$^{(a)}$\  AU means arbitrary unit.
\end{minipage}
\end{table}

\section{Results}

\subsection{Phenotype defined by the epigenetic state of cells}
\label{sec:3.1}

\begin{figure}[htbp]
\centering
\includegraphics[width=10cm]{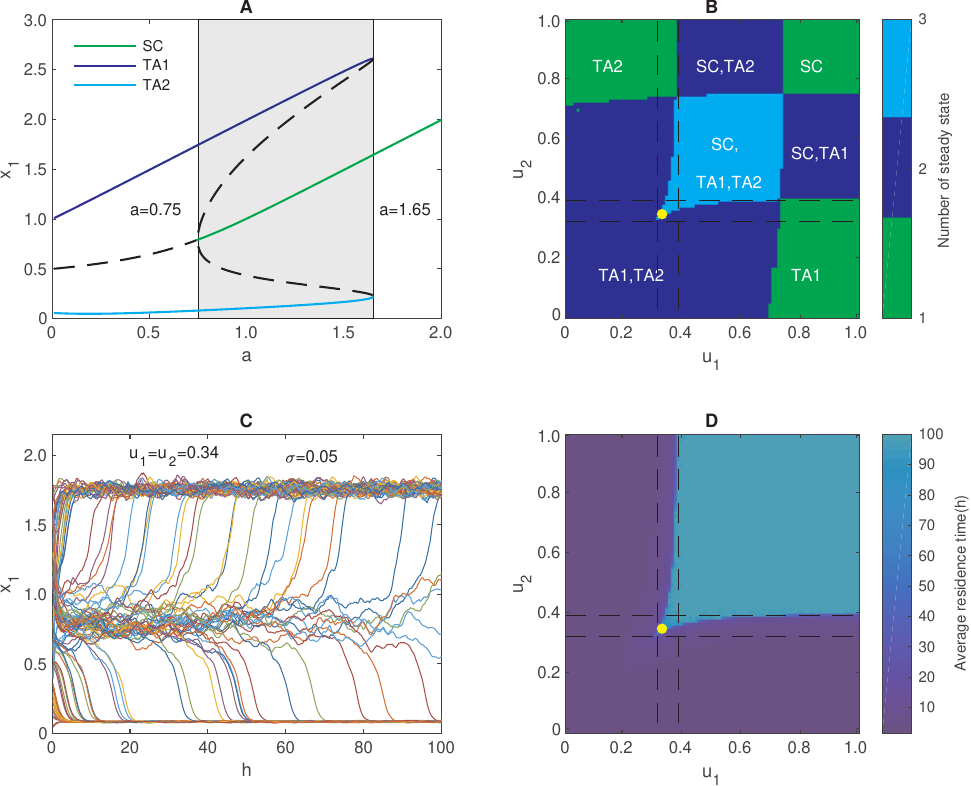}
\caption{Bifurcation analysis of the gene expression dynamics. (A) Dependence of the steady state solution (in $x_1$) on the expression rate $a$ ($a_1 = a_2 = a$). Solid lines represent stable steady states, and dashed lines represent unstable steady states. (B) Dependence of the cell types with epigenetic states $u_1$ and $u_2$.  The color shows the number of steady states. (C) Sample dynamics of cell state transition obtained by solving the stochastic differentiation equation \eqref{eq:1}-\eqref{eq:y_to_a_2}. Here, $u_1 = u_2 = 0.34$  (the yellow dot in B and in D) and $\sigma = 0.05$. (D) Average duration of the SC state with the epigenetic state $(u_1, u_2)$.}
\label{fig:res1_1}
\end{figure}

To quantitatively define the phenotypes of SC and TA cells based on the gene expression state $\bm{x} = (x_1, x_2)$, we performed bifurcation analysis for the ordinary differential equation \eqref{eq:1}. Here, we assumed a symmetric situation so that $a_1 = a_2 = a$ and consider the bifurcation concerning the expression rate $a$. Figure \ref{fig:res1_1}A shows the dependence of the equilibrium state on the parameter $a$.  When $a$ is large ($a > 1.65$), there is a stable state steady with high expressions in both $X_1$ and $X_2$, which is denoted as $(+, +)$. When $a$ decreases ($0.75 < a < 1.65$), in addition to the state $(+, +)$, there are two other stable steady states, one state has high expression in $X_1$ and low expression in $X_2$; the other one is opposite, with low expression in $X_1$ and high expression in $X_2$. We denote these two states as $(++, -)$ and $(-, ++)$, respectively. When $a$ further decreases ($a < 0.75$), the state $(+, +)$ vanishes, and the two states  $(++, -)$ and $(-, ++)$ persist. Biologically, when $a$ decreases from a large ($a > 1.65$) to a small ($a < 0.75$) value, the cell type $(+, +)$ emerge firstly, followed by the coexistence of the three states, and finally, transit to the cell type of either $(++, -)$ or $(-, ++)$. This is akin to the differentiation of precursor cells to either granulocyte/monocyte lineage or erythroid/megakaryocyte lineage described with the PU.1-GATA1 gene circuit of hematopoietic stem cells. Thus, we consider the state $(+, +)$ as stem cells (SC), while the states $(++, -)$ and $(-, ++)$ are downstream transit-amplifying cells (TA1 and TA2).  {Additionally, the states not included in the above steady states are termed transition cells (TC).}  {These arguments define cell types given by equation \eqref{eq:phenotype}}.

Considering the effects of epigenetic modifications and extrinsic noise perturbations, the expression rates $a_i$ are expressed as  {equations} \eqref{eq:y_to_a_2}.  {To begin with, we omitted the extrinsic noise by setting $\sigma_1 = \sigma_2 = 0$}. The dependence of the phenotypes of steady state on the epigenetic state $u_1$ and $u_2$ is shown in Figure \ref{fig:res1_1}B.  Figure \ref{fig:res1_1}B suggests that the three cell types SC, TA1, or TA2 may occur when the epigenetic states of the two genes vary. Specifically, the stem cell state can emerge when $u_1 > 0.4$ and $u_2 > 0.4$. 

Next, we set $\sigma = 0.05$ to introduce the noise perturbation. In this case, the gene expression dynamics are described by the random differential equations \eqref{eq:1}-\eqref{eq:y_to_a_2}. To explore the dynamics of cell-type transition under noise perturbations, given the epigenetic state $u_1$ and $u_2$, we set the initial condition following Figure \ref{fig:res1_1}B and numerically solve equations \eqref{eq:1}-\eqref{eq:y_to_a_2} for $100$h.  Simulations show that TA cells remained unchanged during the simulation. However, for SC with $(u_1, u_2)$ take values near the edge of the SC zone  {($u_1 = u_2 = 0.34$)}, the cell switched to TA cell following random perturbations (Fig. \ref{fig:res1_1}C). Figure \ref{fig:res1_1}D shows the average duration of the SC state with different epigenetic states $(u_1, u_2)$. These results indicate that the definition of cell types of SC, TA1, and TA2 is  {appropriate} following the gene regulation dynamics \eqref{eq:1}-\eqref{eq:y_to_a_2}, and the epigenetic state $(u_1, u_2)$ is important for the phenotype of cells.  

 To further explore the dynamics of cell differentiation induced by noise perturbations in the absence of epigenetic state changes, we investigated the dynamics with the coexistence of SC and TA cells by fixing four sets of epigenetic states: $(u_1, u_2) = (0.5, 0.5), (0.5, 0.6), (0.6, 0.5)$, and $(0.6, 0.6)$. We varied the noise perturbation strength $\sigma$ ranging from $0$ to $1$ and ran the model $100$ times for each combination of epigenetic states and noise perturbation strength, with an initial population of $100$ stem cells and a simulation duration of $1000$h.

\begin{figure}[htbp]
\centering
\includegraphics[width=10cm]{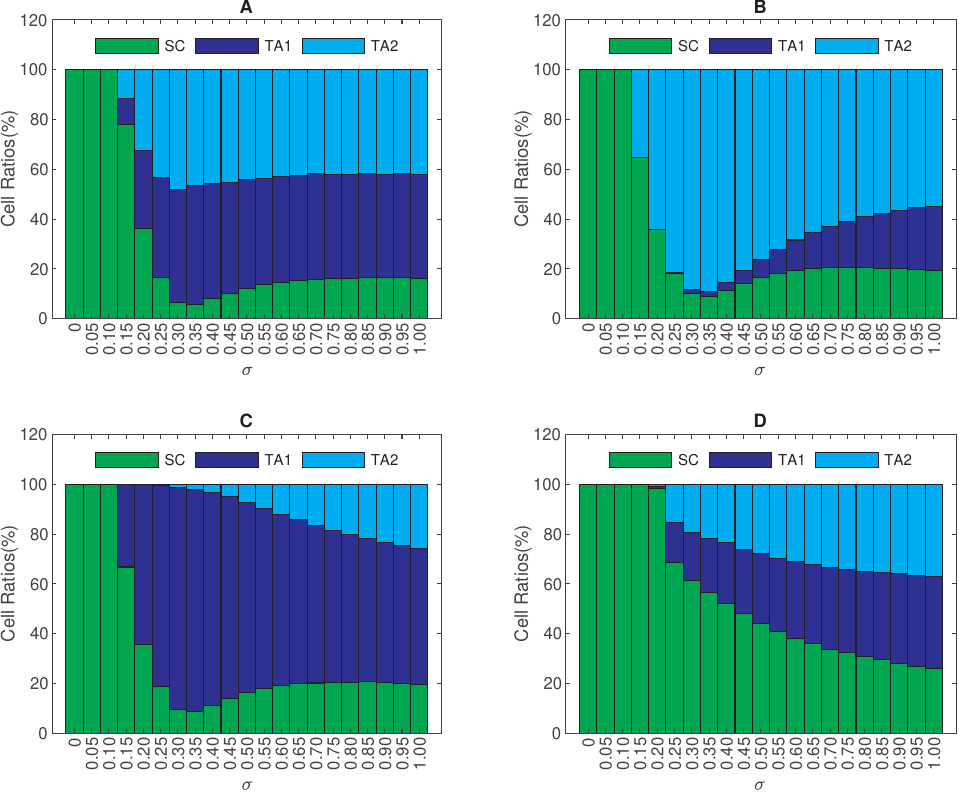}
\caption{The dynamics of cell differentiation induced by noise perturbations in the absence of epigenetic state changes. Figures show the average ratios of SC and TA cells at the end of 100 model runs with different noise perturbation strengths for different sets of epigenetic states: (A) $(u_1, u_2) = (0.5, 0.5)$, (B) $(u_1, u_2) = (0.5, 0.6)$, (C) $(u_1, u_2) = (0.6, 0.5)$, (D) $(u_1, u_2) = (0.6, 0.6)$. Other parameters remained the same as in Table \ref{tab:para}. The statistics are obtained from 100 model runs.}
\label{fig:res1_2}
\end{figure}

Figure \ref{fig:res1_2} shows the average ratios of SC and TA cells with different noise perturbation strengths and epigenetic states at the end of the simulation in $100$ model runs. The results demonstrate that TA cells emerge only when the extrinsic noise strength is larger than a certain threshold. When $(u_1, u_2) = (0.5, 0.5), (0.5, 0.6)$ or $(0.6, 0.5)$, the TA cells appear only when $\sigma > 0.15$, and when $(u_1, u_2) = (0.6, 0.6)$, TA cells appear only when $\sigma > 0.25$. These imply that strong extrinsic noise is required to drive SC differentiation in the absence of epigenetic state changes. Moreover, Figure \ref{fig:res1_2}B and C show that the final fractions of cell types are highly sensitive to the epigenetic state $ (u_1, u_2)$ and the extrinsic noise strength $\sigma$. The interplay between these two factors plays a crucial role in determining the cell fate and type in the multicellular system.

\subsection{Population dynamics of stem cell regeneration and differentiation}

\begin{figure}[htbp]
\centering
\includegraphics[width=10cm]{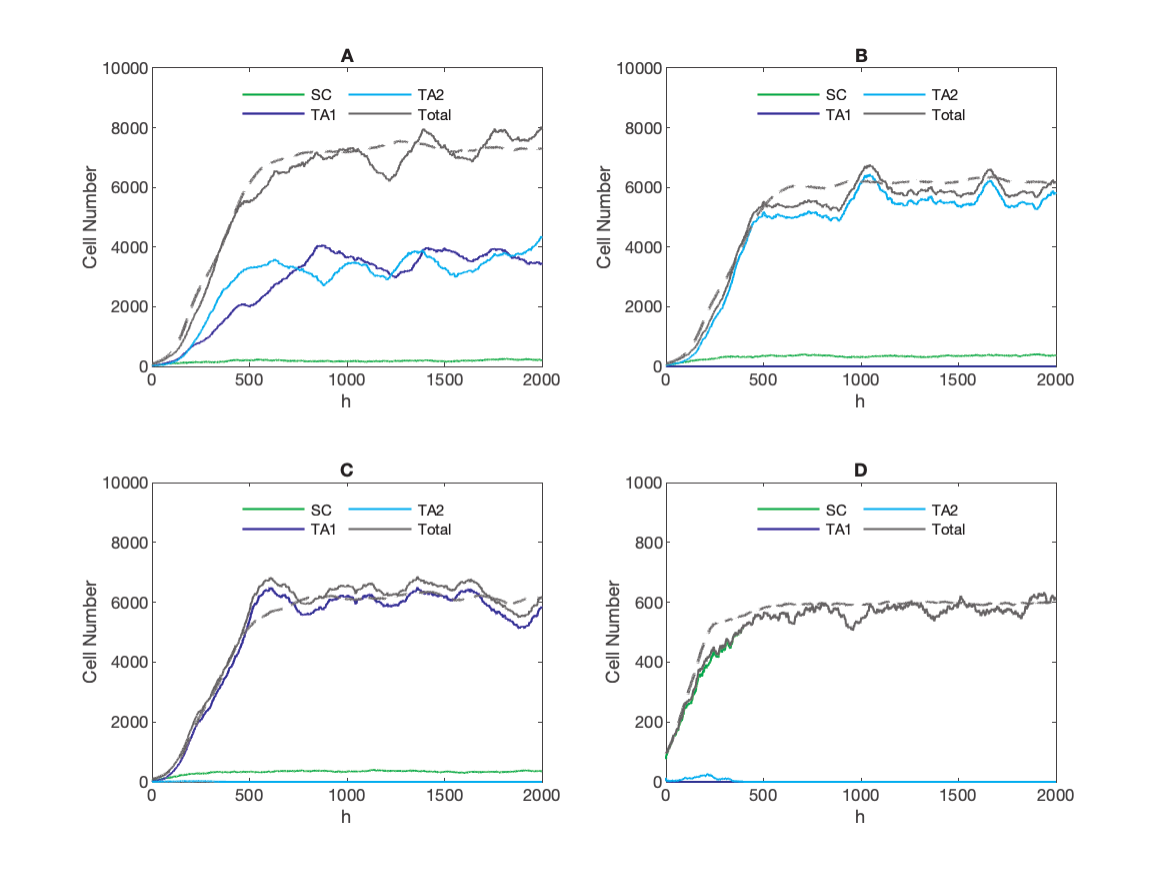}
\caption{Population dynamics of stem cell regeneration and differentiation. Figures show the evolution dynamics of SC and TA cells number with different parameter sets: (A) $(\alpha_1, \alpha_2) = (0.4, 0.4)$, (B) $(\alpha_1, \alpha_2) = (0.4, 1.0)$, (C) $(\alpha_1, \alpha_2) = (1.0, 0.4)$, (D) $(\alpha_1, \alpha_2) = (1.0, 1.0)$. Other parameters remained the same as in Table \ref{tab:para}.  The dashed lines represent the averages of the total cell numbers of 30 model runs.}
\label{fig:res2_1}
\end{figure}

To investigate the population dynamics of stem cell regeneration and differentiation, we considered the full model introduced in Section \ref{model}. Initially, we ran the model simulation with $100$ cells, where the epigenetic state $\bm{u}$ and gene expression state $\bm{x}$ of each cell were randomly distributed over the ranges $0< u_i< 1$ and $0 < x_i < 3$, respectively. To examine the population dynamics with different expression rates of the two genes, we tested four sets of parameters: $(\alpha_1, \alpha_2) = (0.4, 0.4), (0.4, 1.0), (1.0, 0.4)$, and $(1.0, 1.0)$. For each set of parameters,  {we ran the model $30$ times} up to $2000$h  {to reach the stationary equilibrium}. In each case, the cells underwent proliferation, differentiation, and cell death,  {eventually reaching a homeostatic state}. Figure \ref{fig:res2_1} shows the four types of population dynamics for different  {parameter sets}. When $(\alpha_1, \alpha_2) = (0.4, 0.4)$, both SC and TA1 and TA2 cells coexisted,  {indicating} the process of self-renewal of stem cells and differentiation to TA1 and TA2 cells (Fig. \ref{fig:res2_1}A). When $(\alpha_1, \alpha_2) = (0.4, 1.0)$, only SC and TA2 cells  {were presented at homeostasis, indicating} the blockage of differentiation from stem cells to TA1 cells (Fig. \ref{fig:res2_1}B). Similarly, when $(\alpha_1, \alpha_2) = (1.0, 0.4)$,  {TA2 cells were absent} at homeostasis (Fig. \ref{fig:res2_1}C),  {suggesting} the blockage of the differentiation from stem cells to TA2 cells.  When $(\alpha_1, \alpha_2) = (1.0, 1.0)$,  {only stem cells were presented at homeostasis, and no stem cell differentiation events} occur during the simulation (Fig. \ref{fig:res2_1}D). These results indicate that the kinetic parameters of the underlying gene regulation dynamics are crucial for cell phenotypes in homeostasis. In this study, we were interested in the coexistence with SC, TA1, and TA2 cells; thus, we used $(\alpha_1, \alpha_2) = (0.4, 0.4)$ in the following discussions.

\begin{figure}[htbp]
\centering
\includegraphics[width=10cm]{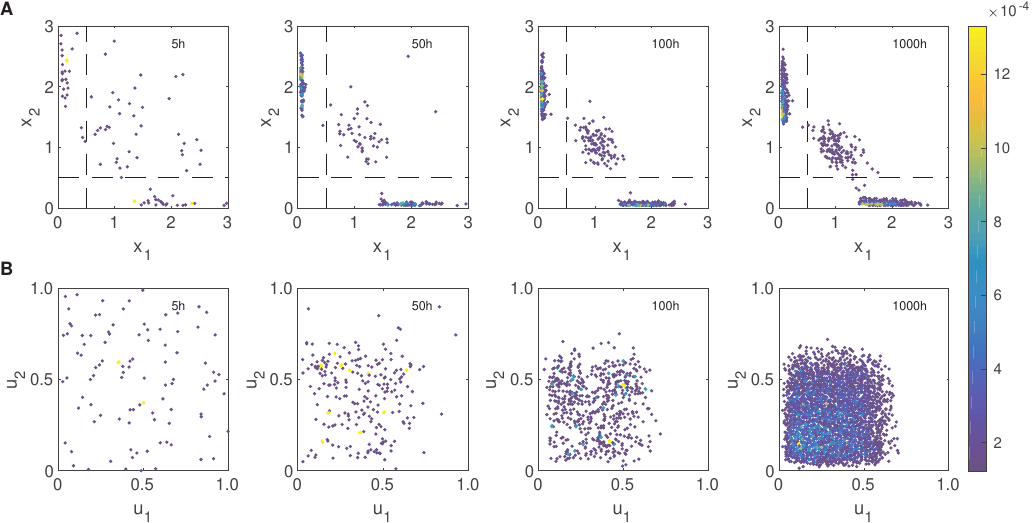}
\caption{The evolution of gene and epigenetic modifications for the population dynamics shown in Figure \ref{fig:res2_1}A. (A) Scatter plots of $(x_1, x_2)$ at different time points ($t =1, 50, 100, 1000$h). (B) Scatter plots of $(u_1, u_2)$ at different points ($t = 5, 50, 100, 1000$h). Parameters are the same as in Table \ref{tab:para}.}
\label{fig:res2_2}
\end{figure}

Next, we further investigated the molecular-level dynamics of individual cells. Figure \ref{fig:res2_2}A shows the scatter plots of $\bm{x} = (x_1, x_2)$ of all cells at different time points  {($t = 5, 50, 100, 1000$h)}. The initially randomly distributed cell states rapidly developed into three clusters  {corresponding} to SC, TA1, and TA2 cells. Accordingly, the epigenetic state $\bm{u}= (u_1, u_2)$ of cells rapidly converged to a steady distribution at homeostasis (Fig. \ref{fig:res2_2}B). Interestingly, despite the continuous distribution of $(u_1, u_2)$, the expression states $(x_1, x_2)$ shown discrete cell types at homeostasis. The bifurcation analysis in Figure \ref{fig:res1_1}  {suggests that} continuous changes in the epigenetic state $\bm{u}$ can lead to transitions of cell types defined by the gene expression state $\bm{x}$. These results indicated that continuous changes in the epigenetic state during stem cell regeneration can lead to discontinuous cell fate decisions,  {providing a mechanism} of stem cell differentiation driven by random inheritance of epigenetic state during cell division.  

\subsection{Dynamics of transdifferentiation and dedifferentiation}

The above simulations  {demonstrated the} differentiation dynamics from stem cells to TA cells,  {a process commonly} occurring during development and tissue homeostasis.  {However, in various biological processes,} transdifferentiation and dedifferentiation are also observed, wherein differentiated cells may lose their phenotype and convert to another cell type or  {revert to} an undifferentiated state\cite{Cobaleda2007,Jopling2011,Collombet2017,Thowfeequ2007}.  {In this context, we have observed that the} random inheritance of epigenetic states can induce the differentiation of stem cells.  {We now explore whether} this mechanism can also lead to transdifferentiation and dedifferentiation. To investigate this, we recorded the cell type changes over a long simulation and counted the number of events of cell type changes in each cell division.  {The average results of 30 model runs are summarized as the transition probability in each cell division in Table \ref{diff-matrix}. }

\begin{table}[htbp]
	\centering
	\caption{ {Cell state transition probabilities ($\%$) from mother to daughter cells at one cell division. The probability values were calculated from the average over 30 model runs.}}
	\label{diff-matrix}
	\begin{tabular}{|c|c|c|c|}
		\hline
		& SC & TA1 & TA2\\
		\hline
		SC & $73.9402$ & $12.9889$ & $13.0709$\\
		\hline
		TA1 & $0.0013$ & $99.9955$ & $0.0032$\\
		\hline
		TA2 & $0.0013$ & $0.0035$ & $99.9952$\\
		\hline
	\end{tabular}
\end{table}

\begin{figure}[htbp]
\centering
\includegraphics[width=10cm]{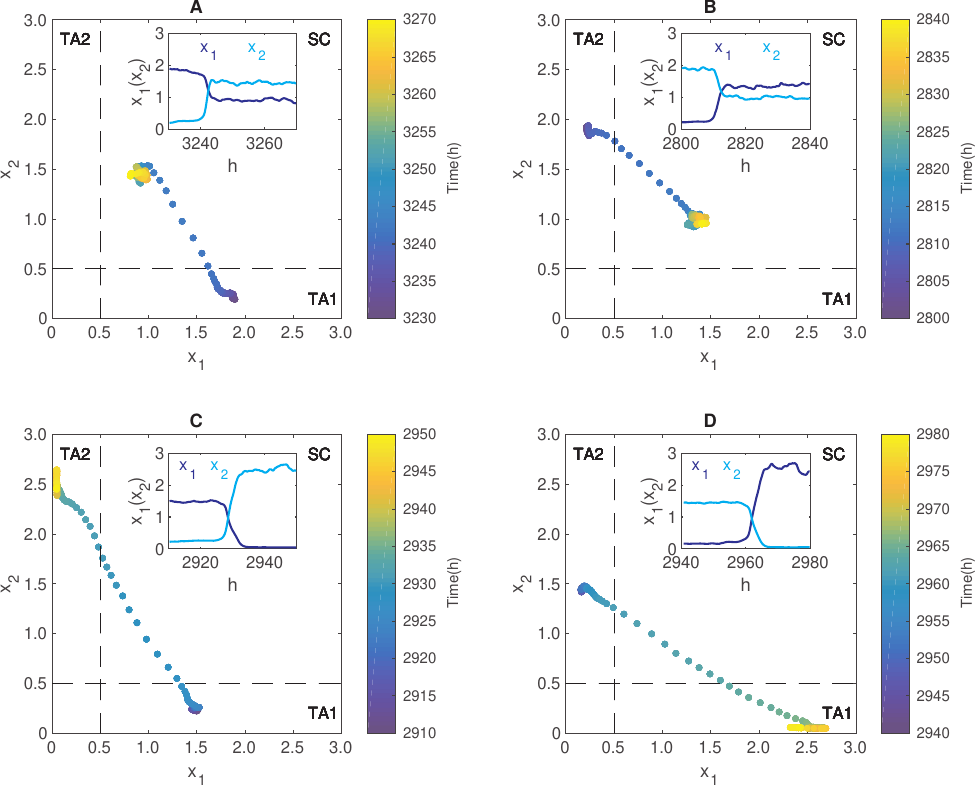}
\caption{Transition trajectories of dedifferentiation, and transdifferentiation in the phase plane. (A) The  {dedifferentiation} trajectory TA1-SC. (B) The  {dedifferentiation} trajectory TA2-SC. (C) The  {transdifferentiation} trajectory TA1-TA2. (D) The  {transdifferentiation} trajectory TA2-TA1. Color bars show the timing in the simulation.  {Insets show the corresponding temporal dynamics $x_1(t)$ and $x_2(t)$. Parameters were the same as in Table \ref{tab:para}.}}
\label{fig:trandyn_2}
\end{figure}

From the results in Table \ref{diff-matrix}, stem cells predominantly underwent self-renewal with a probability of  {73.9402\%, or transit to either TA1 or TA2 cells with probabilities of 12.9889\% or 13.0709\%, respectively, during each cell cycle. However, the TA cells are  {highly inclined} self-renewal, with probabilities exceeding 99.99\%}. Nevertheless,  rare events of transdifferentiation (from TA1 to TA2 or from TA2 to TA1) and dedifferentiation (from TA1 or TA2 to SC) did occur in our simulations,  {albeit with} extremely low frequencies.  {It is important to} note that the transition probabilities depend on the model parameters. In the next section, we discuss how changes in model parameters may affect the frequencies of transdifferentiation and dedifferentiation.  Figure \ref{fig:trandyn_2} shows the transition trajectories of dedifferentiation and transdifferentiation from model simulations in the phase plane of gene expressions. Figure \ref{fig:trandyn_2}C-D shows that the TA states can switch to each other without an intermediate SC state, which suggests evidence of transdifferentiation. 

\subsection{The effects of extrinsic noise and epigenetic state inheritance on homeostasis and cell-type transitions}

 {In this section, we explore} the impact of extrinsic noise and epigenetic state inheritance on the system's homeostasis and the probabilities of cell-type transitions.   {From Figure \ref{fig:res1_2}, cell-type transitions would not happen when the extrinsic noise is weak and in the absence of epigenetic state changes. Moreover, when the noise strength $\sigma$ increases to $\sigma= 0.2$, stem cells may become extinct due to the large differentiation rate (data not shown). Here, we considered the combination of extrinsic noise and epigenetic state changes and assumed a weak extrinsic noise ($\sigma \leq 0.1$).}

\subsubsection{Effect of extrinsic noise}
\begin{figure}[htbp]
\centering
\includegraphics[width=10cm]{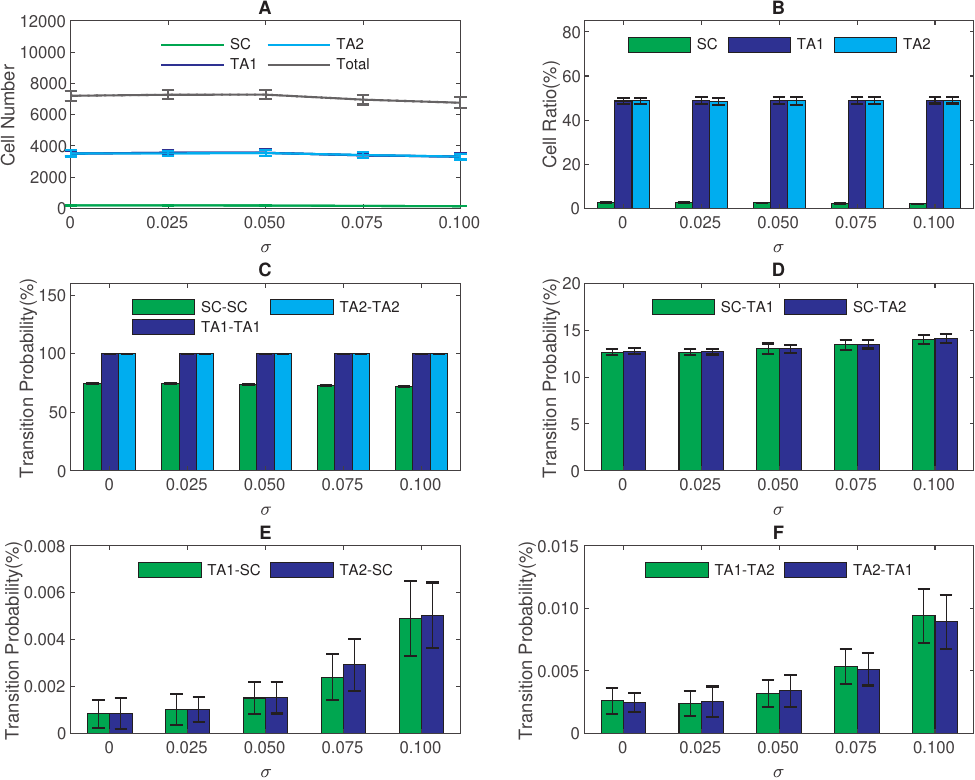}
\caption{The impact of extrinsic noise on homeostasis and cell state transition. (A) Numbers of different cell types at homeostasis.  (B) Ratios of SC and TA cells at homeostasis. (C) The probability of self-renewal in each cell division for different cell types. (D) The probability of differentiation (SC-TA1, SC-TA2) in each cell division. (E) The probability of dedifferentiation (TA1-SC, TA2-SC) in each cell division. (F) The probabilities of transdifferentiation (TA1-TA2, TA2-TA1) in each cell division. All values are calculated from the time scope of  {2000h to 4000h}  in the model simulation. Other parameters were the same as in Table \ref{tab:para}. The statistics are obtained from 30 model runs.}
\label{fig:sigma}
\end{figure}

 We began by varying the noise perturbation strength $\sigma$ from $0$ to $0.1$ and analyzing the homeostasis cell numbers and transition probabilities. For each value of $\sigma$, we conducted model simulation for a time scope of  {4000h so that there are enough data for statistics, and considered the results from 2000h to 4000h for further analysis.} Similar to the previous simulations, the cells developed into SC or TA cells and approached homeostasis. The numbers of different cell types at homeostasis  {was insensitive to changes} in the noise strength(Fig. \ref{fig:sigma}A). The ratios of different types were largely unaffected by the noise strength (Fig. \ref{fig:sigma}B). These findings indicate that changes in extrinsic noise have minimal impact on the system's state at homeostasis. 

 {Next, we investigated the transition probabilities between different cell types under varying noise strengths (Figure \ref{fig:sigma}C-F). The probabilities of self-renewal (SC-SC, TA1-TA1, TA2-TA2) was found to be unaffected by changes in the noise strength (Fig. \ref{fig:sigma}C). The probabilities of differentiation (SC-TA1 and SC-TA2) showed increases with the noise strength $\sigma$ (Fig. \ref{fig:sigma}D); accordingly, the probabilities of dedifferentiation (TA1-SC and TA2-SC) increases with $\sigma$ (Fig. \ref{fig:sigma}E). Likewise, the probabilities of transdifferentiation (TA1-TA2 and TA2-TA1) showed slight increases with $\sigma$ (Fig. \ref{fig:sigma}F). These results suggest that alterations in the extrinsic noise perturbation have minor effects on cell-type transition rates. Therefore, weak extrinsic noise might not be the primary driving force behind cell-type transitions.} 

 {\subsubsection{Effect of epigenetic state inheritance}}

\begin{figure}[htbp]
\centering
\includegraphics[width=10cm]{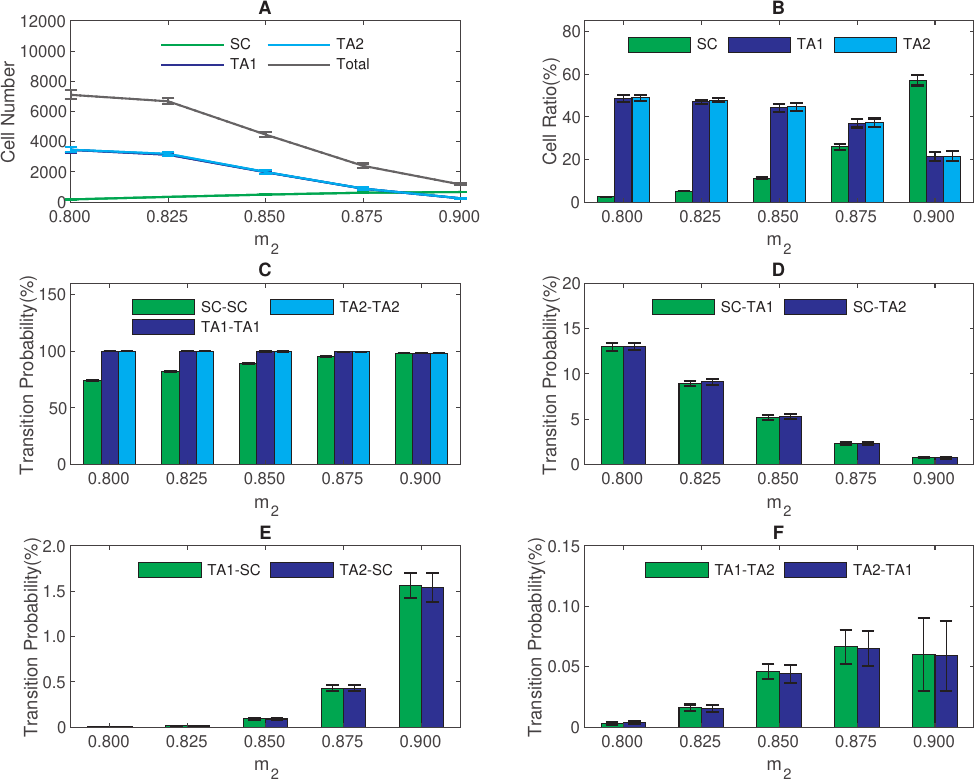}
\caption{The impact of epigenetic state inheritance function on homeostasis and cell state transition. (A) Numbers of different cell types at homeostasis. (B) Ratios of SC and TA cells at homeostasis. (C) The probability of self-renewal in each cell division of different cell types. (D) The probability of differentiation (SC-TA1, SC-TA2) in each cell division. (E) The probability of dedifferentiation (TA1-SC, TA2-SC) in each cell division. (F) The probabilities of transdifferentiation (TA1-TA2, TA2-TA1) in each cell division. All values are calculated from the time scope of 2000h to 4000h in the model simulation. Other parameters were the same as in Table \ref{tab:para}. The statistics are obtained from 30 model runs.}
\label{fig:m2}
\end{figure}

Next, we investigated how random changes in epigenetic modifications may affect the system's homeostasis and cell-type transitions. To this end, we considered the functions $\phi_i(\bm{v})$ in \eqref{eq:epiphi}, which define the inheritance function of epigenetic states. We varied the parameter $m_2$ over the interval $[0.8, 0.9]$ and investigated how the cell numbers and transition probabilities may change with $m_2$. Here, we fixed $\sigma=0.05$, while other parameters remained the same as in Table \ref{tab:para}.  {The simulations revealed that as $m_2$ increased, the numbers of TA cells decreased, while the SC number increased, leading to a decrease in the total cell number} (Fig. \ref{fig:m2}A). Consequently, the ratio of SC increased while the ratio of TA1 and TA2 decreased (Fig. \ref{fig:m2}B). These results demonstrate that alterations in the inheritance functions of epigenetic states significantly impact the system's homeostasis. 

Further examining the probabilities of cell differentiation, transdifferentiation, and dedifferentiation,  {we found that the probabilities} of self-renewal for TA cells were mostly unaffected by changes in $m_2$, while the self-renewal probability of SC increased $m_2$ (Fig. \ref{fig:m2}C). Consequently, the differentiation probabilities (SC-TA1 and SC-TA2) markedly decreased with increasing $m_2$ (Fig. \ref{fig:m2}D), and the dedifferentiation probabilities (TA1-SC and TA2-SC) increased with $m_2$ (Fig. \ref{fig:m2}E). The transdifferentiation probabilities (TA1-TA2 and TA2-TA1)  {showed an initial increase followed by a decrease as $m_2$ was raised}. Biologically, the function $\phi_i(\bm{v})$ represents the expectation of the epigenetic state of daughter cells given the state $\bm{v}$ of the mother cells,  {and $m_2$ can be used to quantify the activities of enzymes that regulate epigenetic modification}. Increasing the parameter $m_2$ upregulates the epigenetic states $u_i$ of daughter cells.  {These results suggest that manipulating epigenetic modifications can be a viable strategy to induce transdifferentiation and dedifferentiation.}

\begin{figure}[htbp]
\centering
\includegraphics[width=10cm]{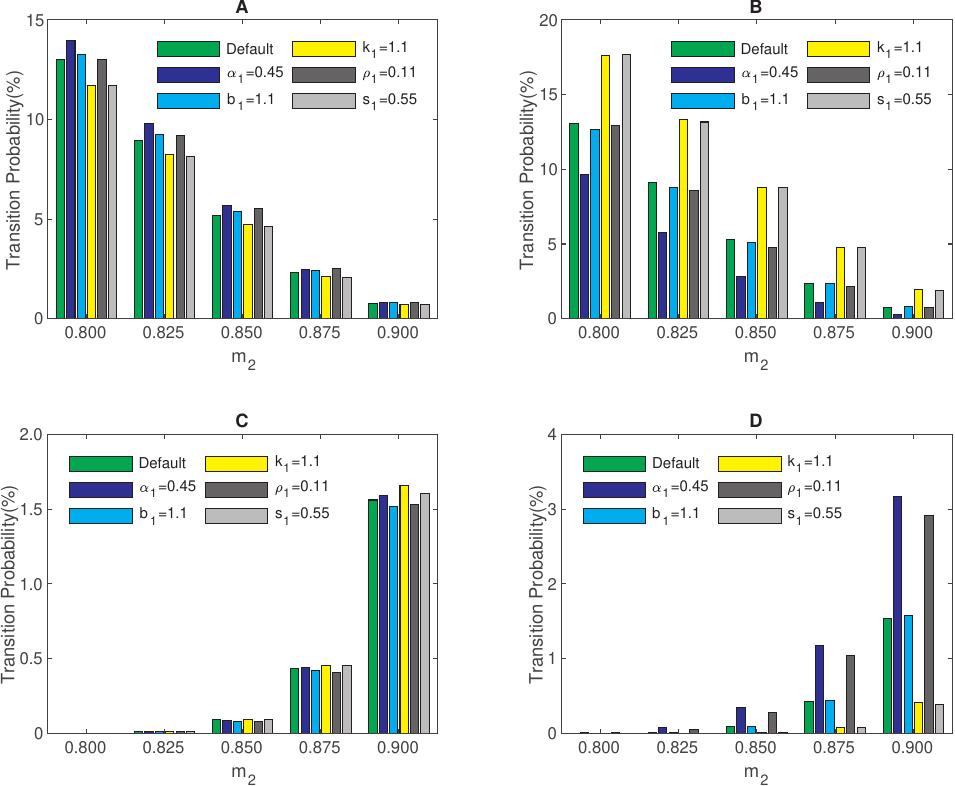}
\caption{ {Transition probability with asymmetry parameters. (A) The transition probability of differentiation SC-TA1 in each cell division. (B) The transition probability of SC-TA2 in each cell division. (C) The probability of dedifferentiation rate TA1-SC in each cell division. (D) The probability of dedifferentiation rate TA2-SC in each cell division. Other parameters were the same as in Table \ref{tab:para}. The statistics were obtained from 30 model runs.}}
\label{fig:asym}
\end{figure}

The aforementioned results were achieved assuming symmetric parameters. In order to explore the impact of asymmetric parameters, we manipulated the gene $X_1$ parameters and conducted model simulations. The transition probabilities in each cell division under distinct parameters are shown in Figure \ref{fig:asym}. The results reveal that asymmetric parameters do not significantly deviate from the principal findings obtained using the assumption of symmetrical parameters. It is worth noting that increases in $k_1$ and $s_1$ correspond to an increase in the differentiation rate SC-TA2 (Fig. \ref{fig:asym}B and a decrease in the dedifferentiation rate TA2-SC (Fig. \ref{fig:asym}D). Conversely, increases in $\alpha_1$ and $\rho_1$ values lead to a reduction in the differentiation rate SC-TA2 (Fig. \ref{fig:asym}B) and an increase in the dedifferentiation rate TA2-SC (Fig. \ref{fig:asym}D).

\subsection{Dynamics of cell reprogramming through the induction of transcription factors}

 {The previous simulations explored the mechanisms of cell transdifferentiation and dedifferentiation induced by extrinsic noise and epigenetic modifications. In this section, we investigate} the dynamics of cell reprogramming through the induction of transcription factors. 

 {We considered introducing an} external transcription factor that enhances the transcription of the self-activation of the $X_1$ gene. Consequently, the equations governing the gene regulation network dynamics are modified as follows:  
 \begin{equation}
\label{eq:TF0}
 \left\{
\begin{aligned}
&\dfrac{\mathrm{d} x_1}{\mathrm{d} t} = (a_1+a_{TF})\left(\rho_1+(1-\rho_1)\dfrac{x_1^{n}}{s_1^n+x_1^{n}}\right)+b_1\dfrac{s_2^{n}}{s_2^n+x_2^{n}}-k_1x_1,\\
&\dfrac{\mathrm{d} x_2}{\mathrm{d} t} = a_2\left(\rho_2+(1-\rho_2)\dfrac{x_2^{n}}{s_2^n+x_2^{n}}\right)+b_2\dfrac{s_1^{n}}{s_1^n+x_1^{n}}-k_2x_2.
\end{aligned}
\right.
\end{equation}
Here,  {$a_{TF}$ represents the augmentation factor} for the activation rate caused by the introduced transcription factor. 

\begin{figure}[htbp]
\centering
\includegraphics[width=10cm]{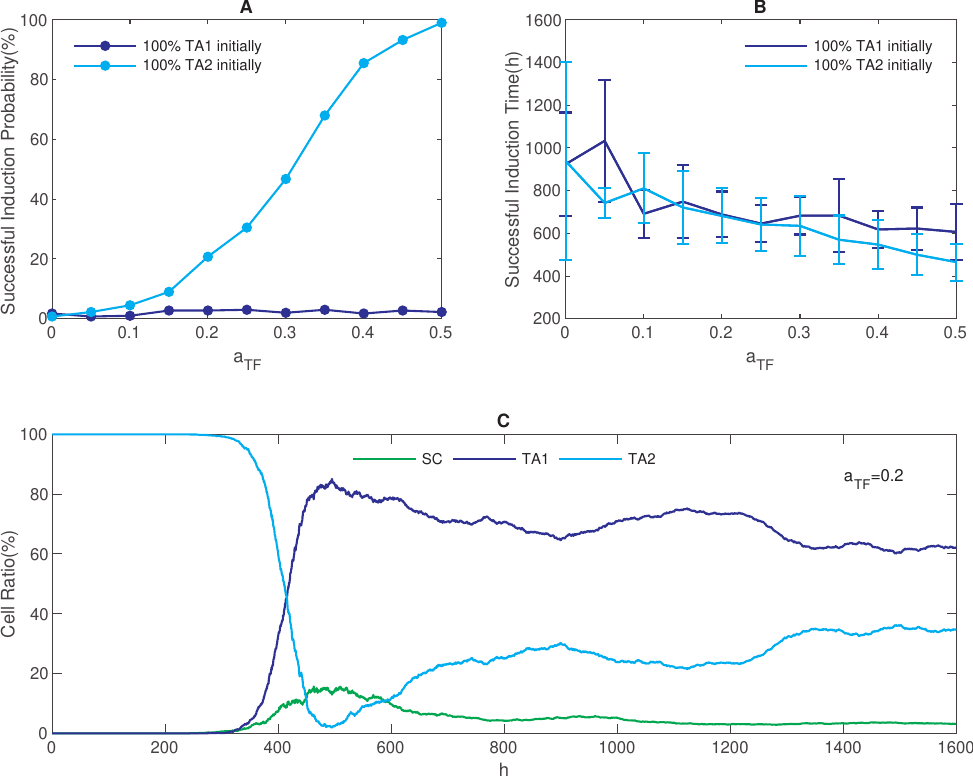}
\caption{Dynamics of cell reprogramming through the induction of transcription factors. (A) The probability of successful stem cell induction. (B) The time required for successful stem cell induction. (C) Time evolution of cell ratios starting from 100\% TA2 cells. Other parameters were the same as in Table \ref{tab:para}. The statistics were obtained from 600 model runs.}
\label{fig:repro}
\end{figure}

To simulate the dynamics of cell reprogramming, we fixed $\sigma=0.05$, $m_2=0.8$, and varied $a_{TF}$ from $0$ to $0.5$, keeping other parameters the same as in Table \ref{tab:para}. In model simulations, the first event is presumably a transition from TA cells to stem cells, and then the system evolves towards the steady state distribution of cell types. We defined successful stem cell induction as the occurrence of 100 stem cells. First, we initialized the system with 500 differentiated cells (either $100\%$ TA1 cells or $100\%$ TA2 cells) and conducted model simulations for 10000h to examine the potential to induce stem cells. When the system initiated with TA1 cells, the probability of successful induction at the end of simulations showed a slight increase and then remained relatively low and stable with varying $a_{TF}$ (Fig. \ref{fig:repro}A). Conversely, when initiated with TA2 cells, the probability of successful induction at the end of simulations increased with $a_{TF}$, and reached a probability of near 100\% when $a_{TF}$ is large enough (Fig. \ref{fig:repro}A). Since the factor $a_{TF}$ only affects the transcription of $X_1$, the introduction of stem cells from TA1 cells is mainly due to the basal dedifferentiation shown in Table \ref{diff-matrix}. Thus, the extra induction probability from TA2 cells comes from the effect of $a_{TF}$. Moreover, we calculate the timing of successful induction in each situation. For the system initiated with TA1 cells, the time required for successful stem cell induction is about 600h with varying $a_{TF}$ (Fig. \ref{fig:repro}B). However, when initiated with TA2 cells,  the time of stem cell induction showed an obvious decrease with the increasing of $a_{TF}$, with the time of 400h when $a_{TF}>0.4$ (Fig. \ref{fig:repro}B). We further show the population dynamics after the induction of $a_{TF}$  (Figure \ref{fig:repro}C). The dynamics reveal that for cells initialized with 100\% TA2 cells, stem cells begin to appear after approximately 300h due to the dedifferentiation of TA2 cells. Subsequently, stem cells differentiate into TA1 and TA2 cells, insulting a cell population approaching homeostasis with all three cell types. 

\begin{figure}[htbp]
\centering
\includegraphics[width=10cm]{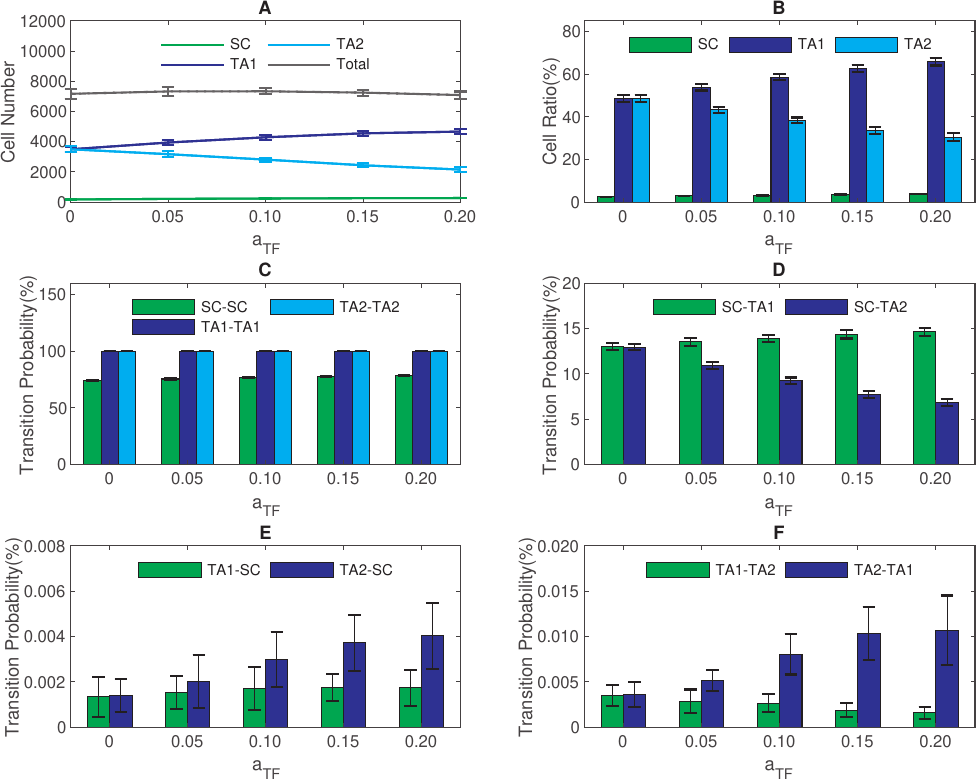}
\caption{The effect of transcription factor introduction on cell number and cell state transition. (A)  {Numbers of different types of cells at homeostasis.} (B) The cell ratios for different cell types over the period from  {2000h to 4000h.} (C)-(F)The cell transition probabilities. Other parameters were the same as in Table \ref{tab:para}. The statistics were obtained from 30 model runs.}
\label{fig:aTF}
\end{figure}

Furthermore, we examined the homeostasis cell numbers and the transition probabilities. We initialized 100 stem cells for statistical simplicity and varied $a_{TF}$ from $0$ to $0.2$. We then conducted model simulations for a time scope of 4000h and analyzed the results from 2000h to 4000h. Consequently, the number of TA1 cells increased with higher values of $a_{TF}$, while the number of TA2 cells decreased (Fig. \ref{fig:aTF}A). The ratios of TA1 cells and stem cells increased with $a_{TF}$, whereas the ratio of TA2 cells decreased with increasing $a_{TF}$ (Fig. \ref{fig:aTF}B). Subsequently, we analyzed the transition probabilities for different values of $a_{TF}$. The self-renewal probability at each cell division remained unaffected by variations in $a_{TF}$, while changes in $a_{TF}$ significantly influenced the probabilities of differentiation, transdifferentiation, and dedifferentiation. Specifically, increasing $a_{TF}$ reduced the differentiation from SC to  {TA2 cells,} promote the dedifferentiation from TA2 cells to SC, and alter the transdifferentiation between TA1 and TA2 cells,  {resulting in an increase in the TA2-TA1 transition probability and a decrease in the TA1-TA2 transition probability} (Fig. \ref{fig:aTF}C-F).

\subsection{Waddington landscape}

 {To gain further insights into the influence of} extrinsic noise and epigenetic state inheritance on cell fate decisions during tissue growth, we  {delve into} the temporal evolution of Waddington landscape based on  {our} simulation results. The numerical scheme described above yields the cell count $N(t, \bm{x})$  {at} time $t$ with state $\bm{x} = (x_1, x_2)$. The total cell number is given by $N(t) = \int N(t, \bm{x}) \mathrm{d} \bm{x}$.  {Consequently, we defined} $f(t, \bm{x})  = N(t, \bm{x})/N(t)$ as the frequency of cells with state $\bm{x}$.  {This enables us to define} the evolution of Waddington's epigenetic landscape as follows:
\begin{equation}
\label{eq:U}
U(t, \bm{x}) = - \log (1 + f(t, \bm{x})),
\end{equation}
where the introduction of the number $1$ prevents issues arising from zero frequency.  

\begin{figure}[htbp]
\centering
\includegraphics[width=10cm]{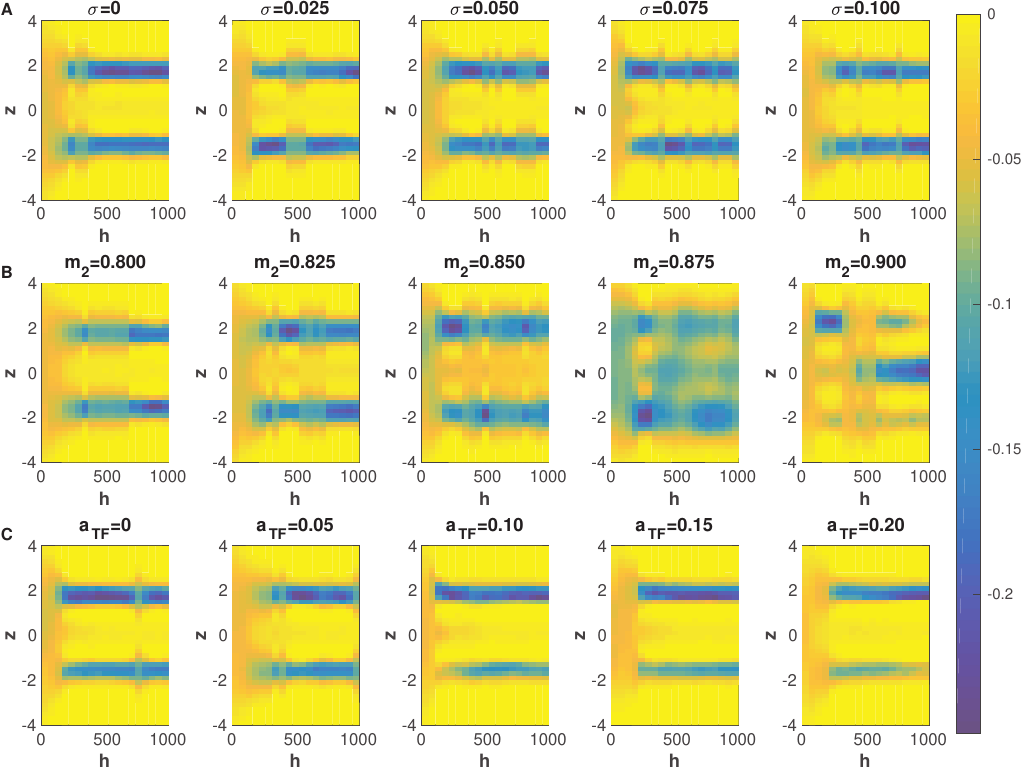}
\caption{Waddington landscape. (A) Temporal evolution of the Waddington landscape with varying extrinsic noises. (B) Temporal evolution of the Waddington landscape with varying epigenetic regulation parameter $m_2$.  (C) Temporal evolution of the Waddington landscape with varying extra factor $a_{TF}$ in cell reprogramming. The Waddington landscapes are visualized as heatmaps for the function $U(t, z)$ with $z = x_1 - x_2$. Other parameters were the same as in Table \ref{tab:para}.}
\label{fig:wad}
\end{figure}

Figure \ref{fig:wad}  {illustrates} landscapes that  {vary with} extrinsic noise strength ($\sigma$), epigenetic regulation parameter ($m_2$), and  {the introduction of an} extra factor ($a_{TF}$) in cell reprogramming. As  {demonstrated} in Figure \ref{fig:wad}A, the landscape  {remains largely unaffected by} changes in the extrinsic noise strength.  {This observation aligns with the above discussion, which revealed that alterations in the extrinsic noise do not impact the homeostatic state of the system} (Fig. \ref{fig:sigma}). However, the landscapes  {exhibit notable changes} as the parameter $m_2$ increases from $0.8$ to $0.9$ (Fig. \ref{fig:wad}B).  {Elevated values of $m_2$} lead to a higher proportion of stem cells in the stationary state, indicating the induction of dedifferentiation through  {alterations} in epigenetic regulation.  {Moreover, the introduction} of the extra factor $a_{TF}$ in \eqref{eq:TF0} disrupts the balance between TA1 and TA2 cells, resulting in  {a greater fraction} of TA1 cells (Fig. \ref{fig:wad}C).  {These findings underscore the substantial role of varying epigenetic regulation and introducing an extra transcription factor in reshaping the Waddington landscape.} 

\section{Discussion}
\label{Discussion}

The intricate regulation of stem cell differentiation and tissue development is a cornerstone challenge in the realms of developmental biology and regenerative medicine. Over time, various mechanisms have been postulated to induce crucial biological processes like cell differentiation, dedifferentiation, and transdifferentiation\cite{Bin2014a,Mojtahedi2016,Stumpf2017,Huang2020}. These mechanisms include stochastic fluctuations, modification to gene regulation networks, the induction of external transcription factors, and the manipulation of small molecules. Yet, while these mechanisms offer valuable insights, they fall short of capturing the inherent progression of cell lineage and the complexity of cellular heterogeneity.

Recent research challenges the prevailing model that assumes changing gene expression within a bistable region is sufficient to alter cell fate\cite{Hoppe2016}. The study by Hoppe et al. demonstrated that random fluctuations of PU.1 and GATA1 expression are insufficient to initiate the cell-fate decision between megakaryocyte/erythroid (MegE) and granulocyte/monocyte (GM) lineages. Long-term absolute protein levels in single differentiating HSCs and their progeny were quantified, which showed that multiple generations are required to induce changes in PU.1 and GATA1 protein levels, and these transcription factors are only executing and reinforcing lineage choice once made\cite{Hoppe2016}. These observations pointed to the significant roles of cell division in guiding cell fate decisions. Additionally, the profound impact of epigenetic regulation on cellular heterogeneity and phenotype transitions has garnered considerable attention\cite{Feinberg2023,Nashun2015,Ashwin2015,Carter2020,Wang2023,Chen2022}.

In our pursuit to quantitatively dissect the dynamics of cell-type transitions driven by epigenetic modifications, we developed a comprehensive hybrid model for stem cell regeneration. This model integrates gene regulation networks, epigenetic state inheritance, and cell regeneration dynamics.  The essence of our model lies in its ability to simulate the biological process of cell population growth and the establishment of homeostasis, characterized by a delicate equilibrium among diverse cell types. Our simulations unravel a mechanism where stochastic inheritance of epigenetic modifications leads to spontaneous switches in cellular phenotypes during cell cycling. We argue that the natural biological process of the cell cycle and stochastic inheritance during cell division can serve as the driving force of cell-type transition.  The interplay between random epigenetic transition and the intrinsic dynamics of gene networks emerges as a linchpin in steering cell-type switches and consolidating cellular stability. Remarkably, our findings underscore how manipulations in epigenetic regulation can wield influence over the epigenetic landscape, thereby augmenting the potential for cell dedifferentiation and transdifferentiation.

The Waddington landscape serves as a core concept in understanding the mechanisms of stem cell differentiation and cellular plasticity\cite{Rajagopal2016,Shakiba2022}. Traditionally, the stochastic dynamics of gene regulatory networks find gene expressions through Langevin equations, or their associated formulations like the master equation or Fokker-Planck equation.  At the heart of this landscape lies the Waddington potential, $U = - \ln (P_{ss})$, intricately linked to the stationary distribution $P_{ss}$ of the master equation or Fokker-Planck equation\cite{Wang2010a,Wang2010b,Li2013a}. This potential is pivotal in revealing the phenotypic outcomes tethered to the underlying regulatory network. However, it falls short of encapsulating the dynamic processes of cell-type transitions during development and tissue growth without considering the cell cycle process. 

 In this context, our study charts a pioneering path by introducing a computational model framework that fuses the mechanics of cell regeneration with the ebbs and flows of epigenetic modifications during cell division. This model allows us to quantitatively map the relationship between alterations in the epigenetic landscape and corresponding phenotypic transitions. Our model simulations breathe life into the concept of temporal potential in a biological context, providing an innovative lens to explore the evolving dynamics of the Waddington landscape throughout tissue growth. With this endeavor, we not only unravel novel insights into the intricate mechanisms governing stem cell differentiation and cell reprogramming but also offers a promising path to enhance the field of regenerative medicine\cite{Jopling2011,Basu2021}.

In conclusion, our research sets forth a novel paradigm by integrating the interplay of gene networks, epigenetic state inheritance, and cell regeneration dynamics. Through detailed computational simulations and biological insights, we shed light on the underlying forces steering cell fate decisions. The holistic understanding gained from our model has the potential to catalyze transformative breakthroughs in regenerative medicine, propelling us closer to unlocking the full regenerative potential of cells and tissues.

\section*{Acknowledgements}
This work is supported by the National Natural Science Foundation of China (No.11831015) and the Science and Technology Project (No.JAT200246) funded by the Education Department of Fujian Province, China.

\section*{References}
\bibliography{refs}

\begin{thebibliography}{10}
\expandafter\ifx\csname url\endcsname\relax
  \def\url#1{\texttt{#1}}\fi
\expandafter\ifx\csname urlprefix\endcsname\relax\def\urlprefix{URL }\fi
\expandafter\ifx\csname href\endcsname\relax
  \def\href#1#2{#2} \def\path#1{#1}\fi

\bibitem{Clevers2015}
H.~Clevers, {What is an adult stem cell?}, Science 350~(6266) (2015)
  1319--1320.
\newblock \href {http://dx.doi.org/10.1126/science.aad7016}
  {\path{doi:10.1126/science.aad7016}}.

\bibitem{Waddington2012}
C.~H. Waddington, {The epigenotype}, Int J Epidemiol 41~(1) (2012) 10--13.
\newblock \href {http://dx.doi.org/10.1093/ije/dyr184}
  {\path{doi:10.1093/ije/dyr184}}.

\bibitem{Ferrell2012}
J.~E. Ferrell, {Bistability, bifurcations, and Waddington's epigenetic
  landscape}, Curr Biol 22~(11) (2012) R458--R466.
\newblock \href {http://dx.doi.org/10.1016/j.cub.2012.03.045}
  {\path{doi:10.1016/j.cub.2012.03.045}}.

\bibitem{Jolly2015a}
M.~K. Jolly, D.~Jia, M.~Boareto, S.~A. Mani, K.~J. Pienta, E.~Ben-Jacob,
  H.~Levine, {Coupling the modules of EMT and stemness: a tunable ‘stemness
  window' model}, Oncotarget 6~(28) (2015) 25161--25174.
\newblock \href {http://dx.doi.org/10.18632/oncotarget.4629}
  {\path{doi:10.18632/oncotarget.4629}}.

\bibitem{Wang2011}
J.~Wang, K.~Zhang, L.~Xu, E.~Wang, {Quantifying the Waddington landscape and
  biological paths for development and differentiation}, Proc Natl Acad Sci USA
  108~(20) (2011) 8257--8262.
\newblock \href {http://dx.doi.org/10.1073/pnas.1017017108}
  {\path{doi:10.1073/pnas.1017017108}}.

\bibitem{Iovino2011}
N.~Iovino, G.~Cavalli, {Rolling ES cells down the Waddington landscape with
  Oct4 and Sox2}, Cell 145~(6) (2011) 815--817.
\newblock \href {http://dx.doi.org/10.1016/j.cell.2011.05.027}
  {\path{doi:10.1016/j.cell.2011.05.027}}.

\bibitem{Zhou2011}
J.~X. Zhou, S.~Huang, {Understanding gene circuits at cell-fate branch points
  for rational cell reprogramming}, Trends Genet 27~(2) (2011) 55--62.
\newblock \href {http://dx.doi.org/10.1016/j.tig.2010.11.002}
  {\path{doi:10.1016/j.tig.2010.11.002}}.

\bibitem{Rommelfanger2021}
M.~K. Rommelfanger, A.~L. MacLean, {A single-cell resolved cell-cell
  communication model explains lineage commitment in hematopoiesis},
  Development 148~(24) (2021) dev199779.
\newblock \href {http://dx.doi.org/10.1242/dev.199779}
  {\path{doi:10.1242/dev.199779}}.

\bibitem{Xu2014}
L.~Xu, K.~Zhang, J.~Wang, {Exploring the mechanisms of differentiation,
  dedifferentiation, reprogramming and transdifferentiation}, PLoS One 9~(8)
  (2014) e105216.
\newblock \href {http://dx.doi.org/10.1371/journal.pone.0105216}
  {\path{doi:10.1371/journal.pone.0105216}}.

\bibitem{Losick2008}
R.~Losick, C.~Desplan, {Stochasticity and cell fate}, Science 320~(5872) (2008)
  65--68.
\newblock \href {http://dx.doi.org/10.1126/science.1147888}
  {\path{doi:10.1126/science.1147888}}.

\bibitem{Cobaleda2007}
C.~Cobaleda, W.~Jochum, M.~Busslinger, {Conversion of mature B cells into T
  cells by dedifferentiation to uncommitted progenitors}, Nature 449~(7161)
  (2007) 473--477.
\newblock \href {http://dx.doi.org/10.1038/nature06159}
  {\path{doi:10.1038/nature06159}}.

\bibitem{Zhou2008}
Q.~Zhou, D.~A. Melton, {Extreme makeover: converting one cell into another},
  Cell Stem Cell 3~(4) (2008) 382--388.
\newblock \href {http://dx.doi.org/10.1016/j.stem.2008.09.015}
  {\path{doi:10.1016/j.stem.2008.09.015}}.

\bibitem{Zhou2008a}
Q.~Zhou, J.~Brown, A.~Kanarek, J.~Rajagopal, D.~A. Melton, {In vivo
  reprogramming of adult pancreatic exocrine cells to $\beta$-cells}, Nature
  455~(7213) (2008) 627--632.
\newblock \href {http://dx.doi.org/10.1038/nature07314}
  {\path{doi:10.1038/nature07314}}.

\bibitem{Yamanaka2012}
S.~Yamanaka, {Induced pluripotent stem cells: past, present, and future}, Cell
  Stem Cell 10~(6) (2012) 678--684.
\newblock \href {http://dx.doi.org/10.1016/j.stem.2012.05.005}
  {\path{doi:10.1016/j.stem.2012.05.005}}.

\bibitem{Amabile2009}
G.~Amabile, A.~Meissner, {Induced pluripotent stem cells: current progress and
  potential for regenerative medicine}, Trends Mol Med 15~(2) (2009) 59--68.
\newblock \href {http://dx.doi.org/10.1016/j.molmed.2008.12.003}
  {\path{doi:10.1016/j.molmed.2008.12.003}}.

\bibitem{Shu2013}
J.~Shu, C.~Wu, Y.~Wu, Z.~Li, S.~Shao, W.~Zhao, X.~Tang, H.~Yang, L.~Shen,
  X.~Zuo, W.~Yang, Y.~Shi, X.~Chi, H.~Zhang, G.~Gao, Y.~Shu, K.~Yuan, W.~He,
  C.~Tang, Y.~Zhao, H.~Deng, {Induction of pluripotency in mouse somatic cells
  with lineage specifiers}, Cell 153~(5) (2013) 963--975.
\newblock \href {http://dx.doi.org/10.1016/j.cell.2013.05.001}
  {\path{doi:10.1016/j.cell.2013.05.001}}.

\bibitem{Angermueller2016}
C.~Angermueller, S.~J. Clark, H.~J. Lee, I.~C. Macaulay, M.~J. Teng, T.~X. Hu,
  F.~Krueger, S.~A. Smallwood, C.~P. Ponting, T.~Voet, G.~Kelsey, O.~Stegle,
  W.~Reik, {Parallel single-cell sequencing links transcriptional and
  epigenetic heterogeneity}, Nat Methods 13~(3) (2016) 229--232.
\newblock \href {http://dx.doi.org/10.1038/nmeth.3728}
  {\path{doi:10.1038/nmeth.3728}}.

\bibitem{Pollen2014}
A.~A. Pollen, T.~J. Nowakowski, J.~Shuga, X.~Wang, A.~A. Leyrat, J.~H. Lui,
  N.~Li, L.~Szpankowski, B.~Fowler, P.~Chen, N.~Ramalingam, G.~Sun, M.~Thu,
  M.~Norris, R.~Lebofsky, D.~Toppani, D.~W. Kemp, M.~Wong, B.~Clerkson, B.~N.
  Jones, S.~Wu, L.~Knutsson, B.~Alvarado, J.~Wang, L.~S. Weaver, A.~P. May,
  R.~C. Jones, M.~A. Unger, A.~R. Kriegstein, J.~A.~A. West, {Low-coverage
  single-cell mRNA sequencing reveals cellular heterogeneity and activated
  signaling pathways in developing cerebral cortex}, Nat Biotechnol 32~(10)
  (2014) 1053--1058.
\newblock \href {http://dx.doi.org/10.1038/nbt.2967}
  {\path{doi:10.1038/nbt.2967}}.

\bibitem{Patel2014}
A.~P. Patel, I.~Tirosh, J.~J. Trombetta, A.~K. Shalek, S.~M. Gillespie,
  H.~Wakimoto, D.~P. Cahill, B.~V. Nahed, W.~T. Curry, R.~L. Martuza, D.~N.
  Louis, O.~Rozenblatt-Rosen, M.~L. Suv{\`{a}}, A.~Regev, B.~E. Bernstein,
  {Single-cell RNA-seq highlights intratumoral heterogeneity in primary
  glioblastoma}, Science 344~(6190) (2014) 1396--1401.
\newblock \href {http://dx.doi.org/10.1126/science.1254257}
  {\path{doi:10.1126/science.1254257}}.

\bibitem{Huang2009}
S.~Huang, {Non-genetic heterogeneity of cells in development: more than just
  noise}, Development 136~(23) (2009) 3853--3862.
\newblock \href {http://dx.doi.org/10.1242/dev.035139}
  {\path{doi:10.1242/dev.035139}}.

\bibitem{Bjorklund2016a}
{\AA}.~K. Bj{\"{o}}rklund, M.~Forkel, S.~Picelli, V.~Konya, J.~Theorell,
  D.~Friberg, R.~Sandberg, J.~Mj{\"{o}}sberg, {The heterogeneity of human
  CD127+ innate lymphoid cells revealed by single-cell RNA sequencing}, Nat
  Immunol 17~(4) (2016) 451--460.
\newblock \href {http://dx.doi.org/10.1038/ni.3368}
  {\path{doi:10.1038/ni.3368}}.

\bibitem{Johnson2015}
M.~B. Johnson, P.~P. Wang, K.~D. Atabay, E.~A. Murphy, R.~N. Doan, J.~L. Hecht,
  C.~A. Walsh, {Single-cell analysis reveals transcriptional heterogeneity of
  neural progenitors in human cortex}, Nat Neurosci 18~(5) (2015) 637--646.
\newblock \href {http://dx.doi.org/10.1038/nn.3980}
  {\path{doi:10.1038/nn.3980}}.

\bibitem{Shalek2013}
A.~K. Shalek, R.~Satija, X.~Adiconis, R.~S. Gertner, J.~T. Gaublomme,
  R.~Raychowdhury, S.~Schwartz, N.~Yosef, C.~Malboeuf, D.~Lu, J.~J. Trombetta,
  D.~Gennert, A.~Gnirke, A.~Goren, N.~Hacohen, J.~Z. Levin, H.~Park, A.~Regev,
  {Single-cell transcriptomics reveals bimodality in expression and splicing in
  immune cells}, Nature 498~(7453) (2013) 236--240.
\newblock \href {http://dx.doi.org/10.1038/nature12172}
  {\path{doi:10.1038/nature12172}}.

\bibitem{Lee2014}
D.-S. Lee, J.-Y. Shin, P.~D. Tonge, M.~C. Puri, S.~Lee, H.~Park, W.-C. Lee,
  S.~M.~I. Hussein, T.~Bleazard, J.-Y. Yun, J.~Kim, M.~Li, N.~Cloonan, D.~Wood,
  J.~L. Clancy, R.~Mosbergen, J.-H. Yi, K.-S. Yang, H.~Kim, H.~Rhee, C.~A.
  Wells, T.~Preiss, S.~M. Grimmond, I.~M. Rogers, A.~Nagy, J.-S. Seo, {An
  epigenomic roadmap to induced pluripotency reveals DNA methylation as a
  reprogramming modulator}, Nat Commun 5~(1) (2014) 5619.
\newblock \href {http://dx.doi.org/10.1038/ncomms6619}
  {\path{doi:10.1038/ncomms6619}}.

\bibitem{Bagci2013}
H.~Bagci, A.~G. Fisher, {DNA demethylation in pluripotency and reprogramming:
  the role of tet proteins and cell division}, Cell Stem Cell 13~(3) (2013)
  265--269.
\newblock \href {http://dx.doi.org/10.1016/j.stem.2013.08.005}
  {\path{doi:10.1016/j.stem.2013.08.005}}.

\bibitem{Zhang2016a}
B.~Zhang, H.~Zheng, B.~Huang, W.~Li, Y.~Xiang, X.~Peng, J.~Ming, X.~Wu,
  Y.~Zhang, Q.~Xu, W.~Liu, X.~Kou, Y.~Zhao, W.~He, C.~Li, B.~Chen, Y.~Li,
  Q.~Wang, J.~Ma, Q.~Yin, K.~Kee, A.~Meng, S.~Gao, F.~Xu, J.~Na, W.~Xie,
  {Allelic reprogramming of the histone modification H3K4me3 in early mammalian
  development}, Nature 537~(7621) (2016) 553--557.
\newblock \href {http://dx.doi.org/10.1038/nature19361}
  {\path{doi:10.1038/nature19361}}.

\bibitem{Feinberg2023}
A.~P. Feinberg, A.~Levchenko, {Epigenetics as a mediator of plasticity in
  cancer}, Science 379~(6632) (2023) eaaw3835.
\newblock \href {http://dx.doi.org/10.1126/science.aaw3835}
  {\path{doi:10.1126/science.aaw3835}}.

\bibitem{Rossetto2012}
D.~Rossetto, N.~Avvakumov, J.~C{\^{o}}t{\'{e}}, {Histone phosphorylation: a
  chromatin modification involved in diverse nuclear events}, Epigenetics
  7~(10) (2012) 1098--1108.
\newblock \href {http://dx.doi.org/10.4161/epi.21975}
  {\path{doi:10.4161/epi.21975}}.

\bibitem{Plass2013}
C.~Plass, S.~M. Pfister, A.~M. Lindroth, O.~Bogatyrova, R.~Claus, P.~Lichter,
  {Mutations in regulators of the epigenome and their connections to global
  chromatin patterns in cancer}, Nat Rev Genet 14~(11) (2013) 765--780.
\newblock \href {http://dx.doi.org/10.1038/nrg3554}
  {\path{doi:10.1038/nrg3554}}.

\bibitem{Probst2009}
A.~V. Probst, E.~Dunleavy, G.~Almouzni, {Epigenetic inheritance during the cell
  cycle}, Nat Rev Mol Cell Biol 10~(3) (2009) 192--206.
\newblock \href {http://dx.doi.org/10.1038/nrm2640}
  {\path{doi:10.1038/nrm2640}}.

\bibitem{Easwaran2014}
H.~Easwaran, H.-C. Tsai, S.~B. Baylin, {Cancer epigenetics: tumor
  heterogeneity, plasticity of stem-like states, and drug resistance}, Mol Cell
  54~(5) (2014) 716--727.
\newblock \href {http://dx.doi.org/10.1016/j.molcel.2014.05.015}
  {\path{doi:10.1016/j.molcel.2014.05.015}}.

\bibitem{Niwa2005}
H.~Niwa, Y.~Toyooka, D.~Shimosato, D.~Strumpf, K.~Takahashi, R.~Yagi,
  J.~Rossant, {Interaction between Oct3/4 and Cdx2 determines trophectoderm
  differentiation}, Cell 123~(5) (2005) 917--929.
\newblock \href {http://dx.doi.org/10.1016/j.cell.2005.08.040}
  {\path{doi:10.1016/j.cell.2005.08.040}}.

\bibitem{Schaffer2010}
A.~E. Schaffer, K.~K. Freude, S.~B. Nelson, M.~Sander, {Nkx6 transcription
  factors and Ptf1a function as antagonistic lineage determinants in
  multipotent pancreatic progenitors}, Dev Cell 18~(6) (2010) 1022--1029.
\newblock \href {http://dx.doi.org/10.1016/j.devcel.2010.05.015}
  {\path{doi:10.1016/j.devcel.2010.05.015}}.

\bibitem{Ralston2005}
A.~Ralston, J.~Rossant, {Genetic regulation of stem cell origins in the mouse
  embryo}, Clin Genet 68~(2) (2005) 106--112.
\newblock \href {http://dx.doi.org/10.1111/j.1399-0004.2005.00478.x}
  {\path{doi:10.1111/j.1399-0004.2005.00478.x}}.

\bibitem{Orkin2008}
S.~H. Orkin, L.~I. Zon, {Hematopoiesis: an evolving paradigm for stem cell
  biology}, Cell 132~(4) (2008) 631--644.
\newblock \href {http://dx.doi.org/10.1016/j.cell.2008.01.025}
  {\path{doi:10.1016/j.cell.2008.01.025}}.

\bibitem{Loh2008}
Y.-H. Loh, J.-H. Ng, H.-H. Ng, {Molecular framework underlying pluripotency},
  Cell Cycle 7~(7) (2008) 885--891.
\newblock \href {http://dx.doi.org/10.4161/cc.7.7.5636}
  {\path{doi:10.4161/cc.7.7.5636}}.

\bibitem{Wang2010a}
J.~Wang, L.~Xu, E.~Wang, S.~Huang, {The potential landscape of genetic circuits
  imposes the arrow of time in stem cell differentiation}, Biophys J 99~(1)
  (2010) 29--39.
\newblock \href {http://dx.doi.org/10.1016/j.bpj.2010.03.058}
  {\path{doi:10.1016/j.bpj.2010.03.058}}.

\bibitem{Lei2009}
J.~Lei, G.~He, H.~Liu, Q.~Nie, {A delay model for noise-induced bi-directional
  switching}, Nonlinearity 22~(12) (2009) 2845--2859.
\newblock \href {http://dx.doi.org/10.1088/0951-7715/22/12/003}
  {\path{doi:10.1088/0951-7715/22/12/003}}.

\bibitem{Austin2006}
D.~W. Austin, M.~S. Allen, J.~M. McCollum, R.~D. Dar, J.~R. Wilgus, G.~S.
  Sayler, N.~F. Samatova, C.~D. Cox, M.~L. Simpson, {Gene network shaping of
  inherent noise spectra}, Nature 439~(7076) (2006) 608--611.
\newblock \href {http://dx.doi.org/10.1038/nature04194}
  {\path{doi:10.1038/nature04194}}.

\bibitem{Lei2014a}
J.~Lei, S.~A. Levin, Q.~Nie, {Mathematical model of adult stem cell
  regeneration with cross-talk between genetic and epigenetic regulation}, Proc
  Natl Acad Sci USA 111~(10) (2014) E880--E887.
\newblock \href {http://dx.doi.org/10.1073/pnas.1324267111}
  {\path{doi:10.1073/pnas.1324267111}}.

\bibitem{Lei2020a}
J.~Lei, {A general mathematical framework for understanding the behavior of
  heterogeneous stem cell regeneration}, J Theor Biol 492 (2020) 110196.
\newblock \href {http://dx.doi.org/10.1016/j.jtbi.2020.110196}
  {\path{doi:10.1016/j.jtbi.2020.110196}}.

\bibitem{Lei2020}
J.~Lei, {Evolutionary dynamics of cancer: from epigenetic regulation to cell
  population dynamics—mathematical model framework, applications, and open
  problems}, Sci China Math 63~(3) (2020) 411--424.
\newblock \href {http://dx.doi.org/10.1007/s11425-019-1629-7}
  {\path{doi:10.1007/s11425-019-1629-7}}.

\bibitem{Moustakas2002}
A.~Moustakas, K.~Pardali, A.~Gaal, C.-H. Heldin, {Mechanisms of TGF-$\beta$
  signaling in regulation of cell growth and differentiation}, Immunol Lett
  82~(1-2) (2002) 85--91.
\newblock \href {http://dx.doi.org/10.1016/S0165-2478(02)00023-8}
  {\path{doi:10.1016/S0165-2478(02)00023-8}}.

\bibitem{Yang2010}
L.~Yang, Y.~Pang, H.~L. Moses, {TGF-$\beta$ and immune cells: an important
  regulatory axis in the tumor microenvironment and progression}, Trends
  Immunol 31~(6) (2010) 220--227.
\newblock \href {http://dx.doi.org/10.1016/j.it.2010.04.002}
  {\path{doi:10.1016/j.it.2010.04.002}}.

\bibitem{Bernard2003}
S.~Bernard, J.~B{\'{e}}lair, M.~C. Mackey, {Oscillations in cyclical
  neutropenia: new evidence based on mathematical modeling}, J Theor Biol
  223~(3) (2003) 283--298.
\newblock \href {http://dx.doi.org/10.1016/S0022-5193(03)00090-0}
  {\path{doi:10.1016/S0022-5193(03)00090-0}}.

\bibitem{Doncic2011}
A.~Doncic, M.~Falleur-Fettig, J.~M. Skotheim, {Distinct interactions select and
  maintain a specific cell fate}, Mol Cell 43~(4) (2011) 528--539.
\newblock \href {http://dx.doi.org/10.1016/j.molcel.2011.06.025}
  {\path{doi:10.1016/j.molcel.2011.06.025}}.

\bibitem{Ginzberg2015}
M.~B. Ginzberg, R.~Kafri, M.~Kirschner, {On being the right (cell) size},
  Science 348~(6236) (2015) 1245075.
\newblock \href {http://dx.doi.org/10.1126/science.1245075}
  {\path{doi:10.1126/science.1245075}}.

\bibitem{Kafri2013}
R.~Kafri, J.~Levy, M.~B. Ginzberg, S.~Oh, G.~Lahav, M.~W. Kirschner, {Dynamics
  extracted from fixed cells reveal feedback linking cell growth to cell
  cycle}, Nature 494~(7438) (2013) 480--483.
\newblock \href {http://dx.doi.org/10.1038/nature11897}
  {\path{doi:10.1038/nature11897}}.

\bibitem{Cadart2018}
C.~Cadart, S.~Monnier, J.~Grilli, P.~J. S{\'{a}}ez, N.~Srivastava, R.~Attia,
  E.~Terriac, B.~Baum, M.~Cosentino-Lagomarsino, M.~Piel, {Size control in
  mammalian cells involves modulation of both growth rate and cell cycle
  duration}, Nat Commun 9~(1) (2018) 3275.
\newblock \href {http://dx.doi.org/10.1038/s41467-018-05393-0}
  {\path{doi:10.1038/s41467-018-05393-0}}.

\bibitem{Zatulovskiy2020}
E.~Zatulovskiy, J.~M. Skotheim, {On the molecular mechanisms regulating animal
  cell size homeostasis}, Trends Genet 36~(5) (2020) 360--372.
\newblock \href {http://dx.doi.org/10.1016/j.tig.2020.01.011}
  {\path{doi:10.1016/j.tig.2020.01.011}}.

\bibitem{Huang2018}
R.~Huang, J.~Lei, {Dynamics of gene expression with positive feedback to
  histone modifications at bivalent domains}, Int J Mod Phys B 32~(07) (2018)
  1850075.
\newblock \href {http://dx.doi.org/10.1142/S0217979218500753}
  {\path{doi:10.1142/S0217979218500753}}.

\bibitem{Huang2019}
R.~Huang, J.~Lei, {Cell-type switches induced by stochastic histone
  modification inheritance}, Discrete Cont Dyn-B 24~(10) (2019) 5601--5619.
\newblock \href {http://dx.doi.org/10.3934/dcdsb.2019074}
  {\path{doi:10.3934/dcdsb.2019074}}.

\bibitem{Jopling2011}
C.~Jopling, S.~Boue, J.~C.~I. Belmonte, {Dedifferentiation,
  transdifferentiation and reprogramming: three routes to regeneration}, Nat
  Rev Mol Cell Biol 12~(2) (2011) 79--89.
\newblock \href {http://dx.doi.org/10.1038/nrm3043}
  {\path{doi:10.1038/nrm3043}}.

\bibitem{Collombet2017}
S.~Collombet, C.~van Oevelen, J.~L. {Sardina Ortega}, W.~Abou-Jaoud{\'{e}},
  B.~{Di Stefano}, M.~Thomas-Chollier, T.~Graf, D.~Thieffry, {Logical modeling
  of lymphoid and myeloid cell specification and transdifferentiation}, Proc
  Natl Acad Sci USA 114~(23) (2017) 5792--5799.
\newblock \href {http://dx.doi.org/10.1073/pnas.1610622114}
  {\path{doi:10.1073/pnas.1610622114}}.

\bibitem{Thowfeequ2007}
S.~Thowfeequ, E.-J. Myatt, D.~Tosh, {Transdifferentiation in developmental
  biology, disease, and in therapy}, Dev Dyn 236~(12) (2007) 3208--3217.
\newblock \href {http://dx.doi.org/10.1002/dvdy.21336}
  {\path{doi:10.1002/dvdy.21336}}.

\bibitem{Bin2014a}
B.~Zhang, P.~G. Wolynes, {Stem cell differentiation as a many-body problem},
  Proc Natl Acad Sci USA 111~(28) (2014) 10185--10190.
\newblock \href {http://dx.doi.org/10.1073/pnas.1408561111}
  {\path{doi:10.1073/pnas.1408561111}}.

\bibitem{Mojtahedi2016}
M.~Mojtahedi, A.~Skupin, J.~Zhou, I.~G. Casta{\~{n}}o, R.~Y.~Y. Leong-Quong,
  H.~Chang, K.~Trachana, A.~Giuliani, S.~Huang, {Cell fate decision as
  high-dimensional critical state transition}, PLoS Biol 14~(12) (2016)
  e2000640.
\newblock \href {http://dx.doi.org/10.1371/journal.pbio.2000640}
  {\path{doi:10.1371/journal.pbio.2000640}}.

\bibitem{Stumpf2017}
P.~S. Stumpf, R.~C. Smith, M.~Lenz, A.~Schuppert, F.-J. M{\"{u}}ller,
  A.~Babtie, T.~E. Chan, M.~P. Stumpf, C.~P. Please, S.~D. Howison, F.~Arai,
  B.~D. MacArthur, {Stem cell differentiation as a non-Markov stochastic
  process}, Cell Syst 5~(3) (2017) 268--282.e7.
\newblock \href {http://dx.doi.org/10.1016/j.cels.2017.08.009}
  {\path{doi:10.1016/j.cels.2017.08.009}}.

\bibitem{Huang2020}
B.~Huang, M.~Lu, M.~Galbraith, H.~Levine, J.~N. Onuchic, D.~Jia, {Decoding the
  mechanisms underlying cell-fate decision-making during stem cell
  differentiation by random circuit perturbation}, J R Soc Interface 17~(169)
  (2020) 20200500.
\newblock \href {http://dx.doi.org/10.1098/rsif.2020.0500}
  {\path{doi:10.1098/rsif.2020.0500}}.

\bibitem{Hoppe2016}
P.~S. Hoppe, M.~Schwarzfischer, D.~Loeffler, K.~D. Kokkaliaris, O.~Hilsenbeck,
  N.~Moritz, M.~Endele, A.~Filipczyk, A.~Gambardella, N.~Ahmed, M.~Etzrodt,
  D.~L. Coutu, M.~A. Rieger, C.~Marr, M.~K. Strasser, B.~Schauberger,
  I.~Burtscher, O.~Ermakova, A.~B{\"{u}}rger, H.~Lickert, C.~Nerlov, F.~J.
  Theis, T.~Schroeder, {Early myeloid lineage choice is not initiated by random
  PU.1 to GATA1 protein ratios}, Nature 535~(7611) (2016) 299--302.
\newblock \href {http://dx.doi.org/10.1038/nature18320}
  {\path{doi:10.1038/nature18320}}.

\bibitem{Nashun2015}
B.~Nashun, P.~W. Hill, P.~Hajkova, {Reprogramming of cell fate: epigenetic
  memory and the erasure of memories past}, EMBO J 34~(10) (2015) 1296--1308.
\newblock \href {http://dx.doi.org/10.15252/embj.201490649}
  {\path{doi:10.15252/embj.201490649}}.

\bibitem{Ashwin2015}
S.~S. Ashwin, M.~Sasai, {Effects of collective histone state dynamics on
  epigenetic landscape and kinetics of cell reprogramming}, Sci Rep 5~(1)
  (2015) 16746.
\newblock \href {http://dx.doi.org/10.1038/srep16746}
  {\path{doi:10.1038/srep16746}}.

\bibitem{Carter2020}
B.~Carter, K.~Zhao, {The epigenetic basis of cellular heterogeneity}, Nat Rev
  Genet 22~(4) (2021) 235--250.
\newblock \href {http://dx.doi.org/10.1038/s41576-020-00300-0}
  {\path{doi:10.1038/s41576-020-00300-0}}.

\bibitem{Wang2023}
X.~Wang, F.~Yu, L.~Ye, {Epigenetic control of mesenchymal stem cells
  orchestrates bone regeneration}, Front Endocrinol (Lausanne) 14~(March)
  (2023) 1--17.
\newblock \href {http://dx.doi.org/10.3389/fendo.2023.1126787}
  {\path{doi:10.3389/fendo.2023.1126787}}.

\bibitem{Chen2022}
C.~Chen, Y.~Gao, W.~Liu, S.~Gao, {Epigenetic regulation of cell fate
  transition: learning from early embryo development and somatic cell
  reprogramming}, Biol Reprod 107~(1) (2022) 183--195.
\newblock \href {http://dx.doi.org/10.1093/biolre/ioac087}
  {\path{doi:10.1093/biolre/ioac087}}.

\bibitem{Rajagopal2016}
J.~Rajagopal, B.~Z. Stanger, {Plasticity in the adult: how should the
  Waddington diagram be applied to regenerating tissues?}, Dev Cell 36~(2)
  (2016) 133--137.
\newblock \href {http://dx.doi.org/10.1016/j.devcel.2015.12.021}
  {\path{doi:10.1016/j.devcel.2015.12.021}}.

\bibitem{Shakiba2022}
N.~Shakiba, C.~Li, J.~Garcia-Ojalvo, K.-H. Cho, K.~Patil, A.~Walczak, Y.-Y.
  Liu, S.~Kuehn, Q.~Nie, A.~Klein, G.~Deco, M.~Kringelbach, S.~Iyer-Biswas,
  {How can Waddington-like landscapes facilitate insights beyond developmental
  biology?}, Cell Syst 13~(1) (2022) 4--9.
\newblock \href {http://dx.doi.org/10.1016/j.cels.2021.12.003}
  {\path{doi:10.1016/j.cels.2021.12.003}}.

\bibitem{Wang2010b}
J.~Wang, C.~Li, E.~Wang, {Potential and flux landscapes quantify the stability
  and robustness of budding yeast cell cycle network}, Proc Natl Acad Sci USA
  107~(18) (2010) 8195--8200.
\newblock \href {http://dx.doi.org/10.1073/pnas.0910331107}
  {\path{doi:10.1073/pnas.0910331107}}.

\bibitem{Li2013a}
C.~Li, J.~Wang, {Quantifying Waddington landscapes and paths of non-adiabatic
  cell fate decisions for differentiation, reprogramming and
  transdifferentiation}, J R Soc Interface 10~(89) (2013) 20130787.
\newblock \href {http://dx.doi.org/10.1098/rsif.2013.0787}
  {\path{doi:10.1098/rsif.2013.0787}}.

\bibitem{Basu2021}
A.~Basu, V.~K. Tiwari, {Epigenetic reprogramming of cell identity: lessons from
  development for regenerative medicine}, Clin Epigenetics 13~(1) (2021) 144.
\newblock \href {http://dx.doi.org/10.1186/s13148-021-01131-4}
  {\path{doi:10.1186/s13148-021-01131-4}}.

\end{thebibliography}

\end{document}